\documentclass[12pt]{article}

\usepackage{graphicx}
\usepackage{amssymb}
\usepackage{amsmath}
\usepackage{pstricks}
\usepackage{amsfonts}
\usepackage{pstricks}
\usepackage{color}
\usepackage{verbatim}
\usepackage{dsfont}

\begin{document}

\begin{titlepage}

\begin{flushright}
\end{flushright}
\vskip 2.5cm

\begin{center}
{\Large \bf Renormalization of Scalar and Yukawa\\\vspace{0.1cm}
Field Theories with Lorentz Violation}
\end{center}

\vspace{1ex}

\begin{center}
{\large Alejandro Ferrero\footnote{{\tt ferrero@physics.sc.edu}}
and Brett Altschul\footnote{{\tt baltschu@physics.sc.edu}}}

\vspace{5mm}
{\sl Department of Physics and Astronomy} \\
{\sl University of South Carolina} \\
{\sl Columbia, SC 29208 USA} \\

\end{center}

\vspace{2.5ex}

\medskip

\centerline{\bf Abstract}

\bigskip

We consider a theory of scalar and spinor fields, interacting through Yukawa and
$\phi^4$ interactions, with Lorentz-violating operators included in the Lagrangian.
We compute the leading quantum corrections in this theory.
The renormalizability of the theory is explicitly shown up to one-loop order. 
In the pure scalar sector, the calculations can be generalized to higher orders
and to include finite terms, because the theory can be solved in terms of its 
Lorentz-invariant version.

\bigskip

\end{titlepage}

\newpage

\section{Introduction}
Lorentz and CPT symmetries seem to be exact in nature. Although they
have been continually confirmed in experiments at relatively low energies ($E<14$ TeV), there is no reason 
to think that they might not be slightly violated at high enough energies, where new physics and
quantum gravitational effects could arise.
The standard approach to study small Lorentz violations (LV) and CPT violations beyond the 
standard model (SM) is the standard model extension (SME)~\cite{ref-reviews,ref-kost2}.
The Lagrange density of the SME contains all possible new operators that can be constructed 
with SM fields using effective field theory. 
In this approach, the fermion and boson fields are coupled to constant background tensors, and these tensors
can be used to parametrize experimental constraints on Lorentz and CPT violations.

Although effective field theory describes what models with Lorentz and CPT violations might look like, it does
not explain what the physics responsible for the LV might be. However, it is known that
any new local operator that violates CPT symmetry will necessarily violate Lorentz symmetry~\cite{ref-greenberg}.
One of the possible scenarios where CPT symmetry can be violated is one where the fundamental
constants are allowed to change. Time variations in the fine structure constant $\alpha$, for instance,
can induce Lorentz and gauge symmetry violations at the quantum level~\cite{ref-lehnert,ref-ferrero1}. LV would also exist
if spacetime is noncommutative~\cite{ref-carroll}. Spacetime discretization
could also be a possible cause of small violations of Lorentz symmetry~\cite{ref-gambini}, as
could spontaneous violation in string theory~\cite{ref-k-sam}.
Whether these scenarios 
exist or not is still an open question; nevertheless, experiments 
searching for any time variation in fundamental constants
or any of the other phenomena that can induce Lorentz symmetry breaking
are of fundamental importance. The absence of these effects in experimental results
is related to tests of Lorentz and CPT symmetries.

Abelian gauge interactions have already been studied in great detail in the SME 
context~\cite{ref-koste-mew,ref-kost5,ref-jackiw1,ref-victoria1,ref-altschul1,ref-altschul2,ref-altschul8,ref-carone}.
Examinations of fermion-photon interactions have led to bounds on many of the parameters contained
in the SME~\cite{ref-stecker,ref-altschul7,ref-altschul7a,ref-altschul7b,ref-hohensee1,ref-altschul20,ref-boquet,ref-altschul22}. 
The Yukawa sector has been rather little studied in the SME context~\cite{ref-Anderson, ref-anber}, in spite 
of its great importance for mass generation in the SM. 
The interactions of scalar fields (such as the Higgs) with fermions are described by Yukawa forces, which can lead to
processes such as $H\rightarrow f+\bar f$. Since the SM
also contains a scalar sector, the study of Lorentz and CPT symmetry
violations in Yukawa interactions is important, and understanding it may be key to comprehending any new
physics that involves LV. With the maximum energies and conditions that are expected to be achieved at the LHC, the production
of Higgs bosons should be possible, which makes this study particularly relevant now. 
However, even if the Higgs particle is not found in the present experiments,
this model can be applied to any kind of interaction between scalar and fermion fields with
possible LV.

Another arena in which Yukawa interactions are important is the study of effective theories containing mesons.
Before quantum chromodynamics was introduced, forces between nucleons were explained by the exchange of pions and other mesons. 
Pions are spin-0 pseudoscalar particles, obeying the Klein Gordon equation and interacting with baryons
through a pseudoscalar Yukawa vertex.
Currently, strong forces are described at its most elemental level by QCD. However, the pion picture can be seen
as a low energy effective theory describing the interactions between nucleons. 

In this paper, the focus of our attention will be quantum corrections rather than tree-level phenomenology; for this purpose, 
we will need to define an appropriate renormalization scheme. Some work on renormalization 
has already been done within the SME context; the main results are in electrodynamics~\cite{ref-kost3,ref-berr},
electroweak interactions~\cite{ref-collad-1}, and non-abelian gauge theories~\cite{ref-collad-2,ref-collad-3}.

This paper is organized as follows: From sections \ref{sec-1} to \ref{ren1} we will focus on a Lorentz-violating 
theory for $N$ scalar fields. We will start by discussing its Lagrangian and Feynman rules in sections
\ref{sec-1}. Quantum corrections for the correlation functions are computed
in section \ref{sec-2}. In section \ref{ren1}, we look at the implications of these results: verifying 
renormalizability, calculating the renormalization group (RG) $\beta$-functions, and showing how the Lorentz-violating
theory may be solved exactly in terms of conventional Lorentz-invariant, $\phi^{\phantom{n}\!\!\!4}$ theory.
In sections \ref{yuk} and \ref{ren2}, the same procedure is applied to a theory with
LV in Yukawa interactions. Section \ref{yuk} introduces its Lagrangian and includes the calculation of the
one-loop corrections to the theory's correlation functions.
The divergences and RG behavior is studied is section \ref{ren2}. Section \ref{concl} summarizes our conclusions.

\section{Lorentz Violation for $N$ Scalar Fields}\label{sec-1}
\subsection{SME Lagrangian}\label{sec-1.1}
A Lorentz-invariant theory with $N$ massive scalar fields, interacting through a $\phi^{\phantom{n}\!\!\!4}$
interaction can be described by the $O(N)$ symmetric Lagrange density
\begin{eqnarray}\label{1.1}
\mathcal{L}=\frac{1}{2}(\partial^\mu \phi_{\phantom{k}\!\!\!i})(\partial_\mu\phi_{\phantom{k}\!\!\!i})
-\frac{1}{2}\mu^2\phi_{\phantom{n}\!\!\!\!i}^{\phantom{k}\!\!\!2}
-\frac{\lambda}{4!}(\phi^2_i)^2,
\end{eqnarray}
where $\phi_i^{\phantom{n}\!\!\!2}=\phi_1^{\phantom{n}\!\!\!2}+\dots+\phi_N^{\phantom{n}\!\!\!2}$.

In the unbroken symmetry phase ($\mu^2>0$), 
the dispersion relation for any excitation of a field $\phi_i$ takes the usual form 
$E=\sqrt{\mathbf{p}^2+\mu^2}$.
Let us now consider adding dimension 3 and 4 Lorentz-violating operators, to give a Lagrange density of the form
\begin{eqnarray}\label{1.2}
\mathcal{L}_K\!\!\!&=&\!\!\!\frac{1}{2}(\partial^\mu \phi_i)(\partial_\mu
\phi_i)+\frac{1}{2}\sum_{i=1}^NK_{\phantom{n}\!\!\!\!\mu\nu}^i\partial^\mu\phi_i\partial^\nu\phi_i
-\frac{1}{2}\mu^2\phi_{\phantom{y}\!\!\!i}^2
+\sum_{i=1}^Nu_i^\beta\phi_i\,\partial_\beta\phi_i\nonumber\\
\!\!\!&&\!\!\!+\,\sum_{j=1}^{N}\phi^{\phantom{n}\!\!\!2}_i\,v_j^\beta\partial_\beta\phi_{j}
-\frac{\lambda}{4!}(\phi^{\phantom{n}\!\!\!2}_i)^2,
\end{eqnarray}
with $|K_{\mu\nu}^i|\ll1$. The terms involving the $u_i^\beta$ and $v_j^\beta$
coefficients in eq.~(\ref{1.2}) are trivial and can be dropped
in a theory with space-time independent coefficients, because they represent total derivatives.
$\mathcal{L}_K$ does not respect Lorentz invariance.
For example, if $K_{\mu\nu}^i=K_{00}^i\delta^0_\mu\delta^0_\nu$,  
the dispersion relation for $\phi_i$ takes the form $E=\sqrt{\mathbf{p}^2-K_{00}^i(p^0)^2+\mu^2}$, 
which is clearly not boost invariant. $\mathcal {L}_K$ is, however, invariant under CPT transformations.

The dimensionless coefficients $K_{\mu\nu}^i$ are
expected to be very small (because Lorentz invariance is at least
approximately valid). The inclusion of higher-dimensional operators
could effectively make them depend on the momentum carried
by the fields, generating additional new effects. However, we will assume them
to be constant and, moreover, equal for each field ($K_{\mu\nu}^i=K_{\mu\nu}$),
preserving the $O(N)$ symmetry.
Their constancy relies on the inference that momentum-dependent terms would
arise at an energy scale far beyond the one of interest.
The symmetry condition $K_{\mu\nu}=K_{\nu\mu}$ is also implied by the structure
of the $K_{\mu\nu}\partial^\mu\phi_i\partial^\nu\phi_i$ term.

\subsection{Feynman Rules for the Renormalized Theory}\label{sec-1.2}
The Lagrangian defined in eq.~(\ref{1.2}) demands the introduction
of new Feynman rules that account for the effects of the coefficients $K_{\mu\nu}$.
Following a perturbative approach, the Lagrangian density will be written as 
$\mathcal{L}_K=\mathcal{L}_0+\mathcal{L}_{int}$, where
\begin{eqnarray}\label{1.2.1}
\mathcal{L}_0=\frac{1}{2}(\partial^\mu \phi_i)(\partial_\mu
\phi_i)+\frac{1}{2}K_{\mu\nu}\partial^\mu\phi_i\partial^\nu\phi_i-\frac{1}{2}\mu^2\phi_i^2
\end{eqnarray}
and
\begin{eqnarray}\label{1.2.2}
\mathcal{L}_{int}=-\frac{\lambda}{4!}(\phi_i^2)^2=-\frac{\lambda}{4!}
\left(\sum_i^N\phi_i^4+2\sum_{i\neq j}^N\phi_i^2\phi_j^2\right).
\end{eqnarray}

Note that no Lorentz-violating modification of $\mathcal{L}_{int}$ is possible without including
operators of higher dimension.
In momentum space, the free Feynman propagator derived from eq.~(\ref{1.2.1}) is
\begin{eqnarray}\label{1.2.3}
D^{ij}_F(p)=D_F(p)\delta^{ij}=\frac{i\,\delta^{ij}}{p^2+K_{\mu\nu}p^\mu p^\nu-\mu^2+i\epsilon}\,.
\end{eqnarray}
Although this is the exact propagator for the free theory,
loop calculations will be
very difficult to perform if the full expression is used. 
Taking advantage of the fact that the coefficients $K_{\mu\nu}$ are
small, we expand eq.~(\ref{1.2.3}) as
\begin{eqnarray}\label{b4}
D_F(p)=\frac{i}{p^2-\mu^2+i\epsilon}\left[1-\frac{K_{\mu\nu}p^\mu
p^\nu}{p^2-\mu^2}+\frac{K_{\mu\nu}K_{\rho\sigma}p^\mu p^\nu p^\rho
p^\sigma}{(p^2-\mu^2)^2}+\cdots\right].
\end{eqnarray}
The free propagator becomes an infinite sum. The inclusion of the first two corrections 
should be an excellent approximation, but the effect of all higher order contributions will also
be included in section \ref{ren1}. (where we also show how using the Lorentz-violating propagator
given by eq.~(\ref{1.2.3}) is equivalent to transforming the action, by means of the introduction of the
Jacobian induced by the matrix $K$.) Using a diagrammatic representation, the 
free propagator for the field $\phi_i$ will be represented by
\begin{center}
\includegraphics[scale=1]{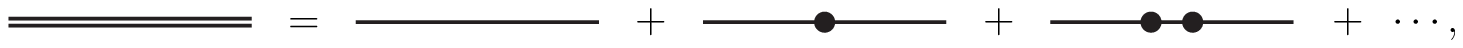}
\end{center}
where each black dot represents a Lorentz-violating $K$ 
insertion.

\begin{figure}
\centering
\includegraphics[scale=1]{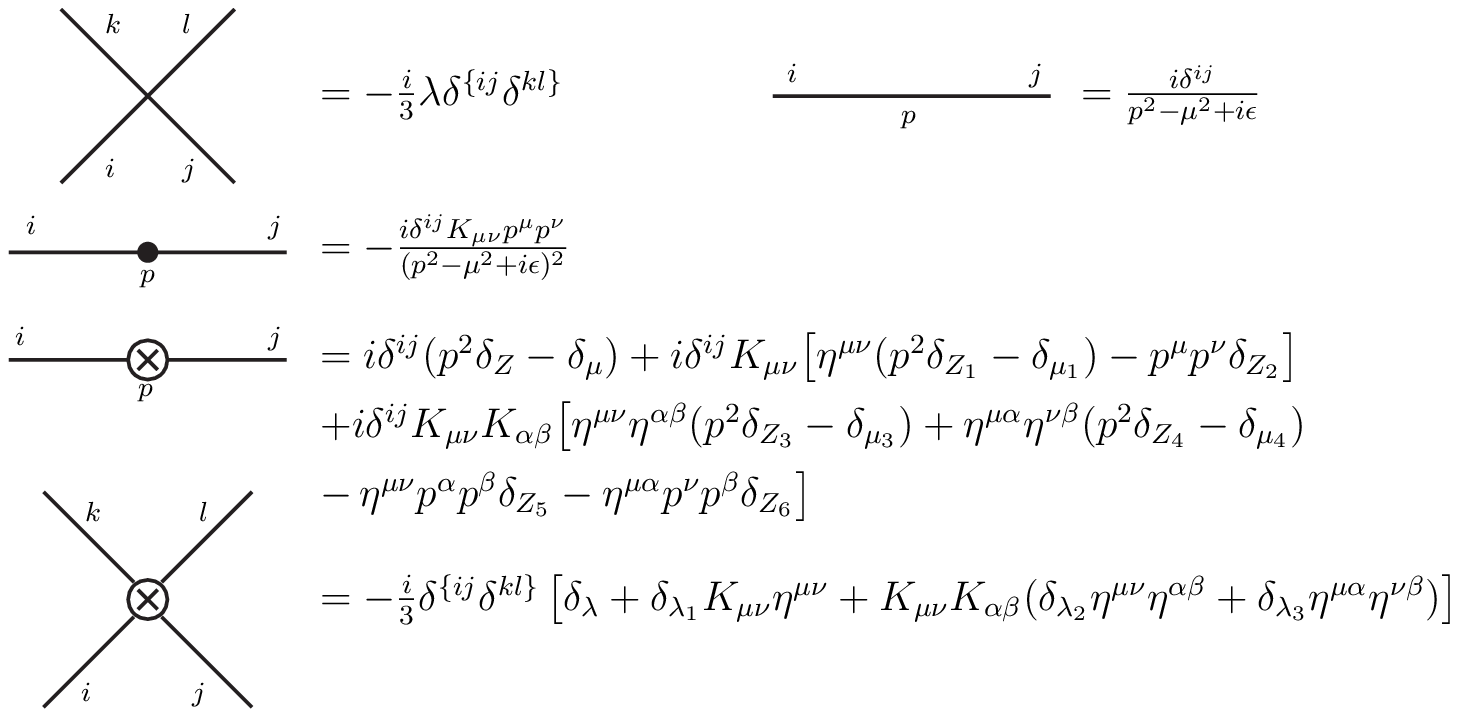}
\caption{Feynman rules for $N$ scalar fields with LV. We
defined $\delta^{\{ij}\delta^{kl\}}\equiv\delta^{ij}\delta^{kl}
+\delta^{ik}\delta^{jl}+\delta^{il}\delta^{jk}$.}
\label{fig-f.rules}
\end{figure}

The vertex is not directly affected by the introduction of $K$.
However, a $K$ dependence arises at the quantum level, because
the renormalized vertex contains internal propagators.
The Feynman rules for this theory are summarized in
fig.~\ref{fig-f.rules}, where the set of counterterms required to renormalize 
the theory is also introduced. 

\section{Corrections to Scalar Correlation Functions}\label{sec-2}
\subsection{The Vertex}\label{sec-2.1}
\begin{figure}
\centering
\includegraphics[scale=1]{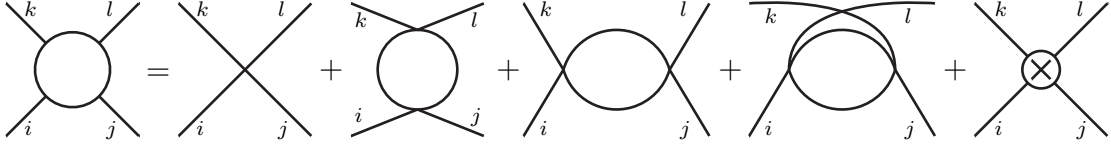}
\caption{Lorentz-invariant one-loop corrections to the scalar four-point function;
the last diagram represents the counterterm.}
\label{fig-2.1a}
\end{figure}

In order to find the corrections to the four-point function, we will include
all the one-loop contributions, which give $\mathcal O(\lambda^2)$ corrections. 
The Lorentz-invariant diagrams that contribute---up to $\mathcal{O}(\lambda^2)$---are 
shown in fig. \ref{fig-2.1a}. 
We begin with the Lorentz-invariant result, which arises with the use of the
propagator $D^{0}_F(p)=\frac{i}{p^2-\mu^2+i\epsilon}$. In terms of the symmetric
sum of indices---$\delta^{\{ab}\delta^{cd\}}=\delta^{ab}\delta^{cd}
+\delta^{ac}\delta^{bd}+\delta^{ad}\delta^{bc}$---this contribution can be written as
\begin{eqnarray}\label{2.1.1}
i\mathcal{M}^{ijkl}_0\!\!\!&=&\!\!\!-\frac{i\lambda}{3}\delta^{\{ij}\delta^{kl\}}
+\left(\frac{-i\lambda}{3}\right)^{\!2}\!i\big[M^{ijkl}V(s)+M^{ikjl}V(t)
+M^{iljk}V(u)\big]-\frac{i}{3}\delta^{\{ij}\delta^{kl\}}\delta_\lambda,\nonumber\\
\end{eqnarray}
where $s,t$ and $u$ denote the usual Mandelstam variables. Contracting the flavor indices
on any of the loop diagrams shown in fig.~\ref{fig-2.1a}, we find
\begin{eqnarray}
M^{ijkl}=(N+2)\delta^{ij}\delta^{kl}+2\delta^{\{ij}\delta^{kl\}}.
\end{eqnarray}
Since $V(p^2)$ includes a divergent contribution, some renormalization
scheme must be introduced. We will use dimensional regularization, where the spacetime
dimension $d$ is $d=4-\epsilon$. The limit $\epsilon\rightarrow0$ produces the final results, 
as usual. Using the Feynman rules,
and denoting by $p$ the momentum transfer, we find
\begin{eqnarray}\label{2.1.2}
V(p^2)=-\frac{1}{2(4\pi)^2}\int_0^1dx\frac{\Gamma(\epsilon/2)}{\Delta^{\epsilon/2}},
\end{eqnarray}
where $\Delta\equiv \mu^2-x(1-x)p^2$.

When we include the corrections arising from the Lorentz-violating terms, the
function $V$ will be modified. For the one-loop case, we will denote
by $V^{(i)}$ the contribution that arises when $i$ insertions of $K$ are introduced 
on the internal lines. $K$ insertions on the external lines
are not interesting, because those diagrams can be reduced by amputating the
Lorentz-violating external legs.
When one $K$ insertion is introduced on one internal propagator, we find a correction
\begin{eqnarray}\label{2.1.2a}
V^{(1)}=\frac{K_{\mu\nu}}{(4\pi)^2}
\int_0^1dx\frac{(1-x)}{\Delta^{\epsilon/2}}\left[\frac{\eta^{\mu\nu}\Gamma(\epsilon/2)}{2}-\frac{x^2p^\mu p^\nu}{\Delta}\right]
\end{eqnarray}
to $V$. Two $K$ insertions can be made in two different ways,
both on the same internal line or one on each line. Combined together, they give 
\begin{eqnarray}\label{2.1.3}
V^{\,(2)}=-\frac{K_{\mu\nu}K_{\rho\sigma}}{2(4\pi)^2}\!\!\!\int_0^{1}\!dx
\frac{(1-x)}{\Delta^{\epsilon/2}}
\Bigg[\frac{\Gamma(\epsilon/2)}{4}\eta^{\{\mu\nu}\eta^{\rho\sigma\}}
-\frac{x}{2\Delta}\eta^{\mu\nu}p^{\rho}p^{\sigma}+\frac{x^3(1-x)}{\Delta^2}p^\mu p^\nu p^{\rho} p^\sigma\Bigg].
\end{eqnarray}
The two-loop corrections were also computed. 
Since the results are rather complicated, they are not shown, but
their contributions to the $\beta$-function will be included.

\subsection{The Scalar Propagator}\label{sec-2.2}
\begin{figure}
\centering
\includegraphics[scale=1]{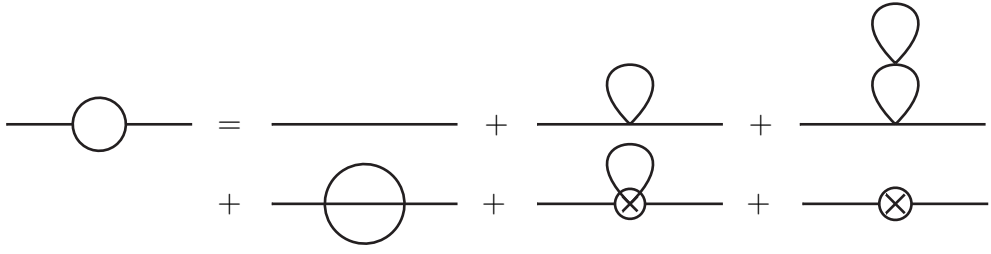}
\caption{Renormalizable $\mathcal O(\lambda^2)$ corrections to the scalar propagator.}
\label{fig-2.2a}
\end{figure}

As is well known, the lowest-order correction to the
scalar propagator in $\phi^4$ theory does not have any momentum dependence. 
Therefore, to study the effects of $K_{\mu\nu}$ on the field strength
renormalization, we must consider two-loop contributions.
In the Lorentz-invariant case, the lowest order corrections---up to two-loop order---to
the scalar propagator are represented by 
the diagrams in fig.~\ref{fig-2.2a}, which can be written as
\begin{eqnarray}\label{2.2.1}
\bar D^{ij}_F(p)\!\!\!&=&\!\!\!\delta^{ij}D_F^0(p)+\frac{-i\lambda}{3}\delta^{ij}D_F^0(p)Z_1^{(0)}(p)D_F^0(p)
+\left(\frac{-i\lambda}{3}\right)^2\delta^{ij}D_F^0(p)iZ_2^{(0)}(p)D_F^0(p)\nonumber\\
\!\!\!&+&\!\!\!D_F^0(p)i\delta^{ij}(p^2\delta_Z-\delta_m)D_F^0(p).
\end{eqnarray}
The loop contribution of $\mathcal O(\lambda^n K^m)$ is denoted $Z_n^{(m)}$.
The tadpole diagram, which is the lowest correction---but momentum-independent---gives the usual infinite contribution
at $O(K^0)$
\begin{eqnarray}\label{2.2.2}
Z_1^{(0)}=\frac{(N+2)}{2(4\pi)^{2}}\frac{\Gamma(1-d/2)}{(\mu^2)^{1-d/2}}.
\end{eqnarray}
The inclusion of Lorentz-violating terms in the tadpole also gives
infinite results. When one or two $K$ insertions appear
inserted on the internal line we find, respectively,
\begin{eqnarray}\label{2.2.3}
Z_1^{(1)}\!\!\!&=&\!\!\!-\frac{(N+2)}{4(4\pi)^{2}}\frac{\Gamma(1-d/2)}{(\mu^2)^{1-d/2}}K_{\mu\nu}\eta^{\mu\nu},\\
Z_1^{(2)}\!\!\!&=&\!\!\!\frac{(N+2)}{16(4\pi)^{2}}
\frac{\Gamma(1-d/2)}{(\mu^2)^{1-d/2}}K_{\mu\nu}K_{\rho\sigma}\eta^{\{\mu\nu}\eta^{\rho\sigma\}}.\,\,\,\,\,
\end{eqnarray}

The first correction to the field strength renormalization comes from
the two-loop sunset diagram---the fourth one shown in fig.~\ref{fig-2.2a}.
The other diagrams do not depend on the momentum, so they can totally be absorbed in the 
mass counterterm. The Lorentz-invariant value of the sunset diagram
is given by the integral
\begin{eqnarray}\label{2.2.4}
\delta^{ij}Z_2^{(0)}=-\frac{i}{3!}\delta^{\{ik}\delta^{lm\}}\delta^{\{op}\delta^{nj\}}
\int\frac{d^dk_1}{(2\pi)^d}\frac{d^dk_2}{(2\pi)^d}\frac{i\delta^{mn}}{(p-k_1-k_2)^2-\mu^2}
\frac{i\delta^{ko}}{k_1^2-\mu^2}\frac{i\delta^{lp}}{k_2^2-\mu^2},
\end{eqnarray}
which becomes after some algebra 
\begin{eqnarray}\label{2.2.5}
Z_2^{(0)}=\frac{(N+2)}{2(4\pi)^4}\int_0^1dx
\int_0^{1-x}dy\frac{\Gamma(-1+\epsilon)}{g^{\,2-\epsilon/2}}(\Delta_1)^{1-\epsilon},
\end{eqnarray}
where $g\equiv x^2-x+xy-y+y^2$, $\Delta_1\equiv \frac{yf}{g}p^2-\mu^2$, and
$f\equiv x^2-x+xy$.
To find the contributions from the Lorentz-violating terms, we introduce a $K$ insertion
on any of the three internal lines. Each such insertion gives the same contribution; thus the total 
result is
\begin{eqnarray}\label{2.2.6}
Z_2^{(1)}=\frac{3(N+2)K_{\mu\nu}}{2(4\pi)^4}\!\int_0^1\!\!dx\!\int_0^{1-x}\!\!\!dy
\frac{y\,(\Delta_1)^{-\epsilon}}{g^{\,3-\epsilon/2}}\!
\left[\frac{\Gamma(-1+\epsilon)}{2}(1-y)\eta^{\mu\nu}\Delta_1-\Gamma(\epsilon)\frac{f^2}{g}p^\mu p^\nu\right]\!.
\end{eqnarray}
When two ${K}$ insertions are introduced in the internal lines, there are two different possibilities.
As in the four-point function case, the Lorentz violating insertions can be introduced on the same line or on different ones. 
Each case has a total of three possibilities, and all of them add up to make
\begin{eqnarray}\label{2.2.7}
Z_2^{(2)}\!\!\!&=&\!\!\!\frac{3(N+2)}{2(4\pi)^4}K_{\mu\nu}K_{\rho\sigma}\int_0^1\!dx\int_0^{1-x}\!\!dy
\frac{y(\Delta_1)^{-\epsilon}}{g^{\,4-\epsilon/2}}
\Bigg\{\frac{\Gamma(-1+\epsilon)}{4}\,\big[y(1-y)^2+2(1-x-y)x^2\big]\nonumber\\
\!\!\!&&\!\!\!\times(\eta^{\mu\nu}\eta^{\rho\sigma}+\eta^{\mu\rho}\eta^{\nu\sigma})\Delta_1
+\frac{\Gamma(\epsilon)}{2}x^2(1-x)(1-x-y)\,\eta^{\mu\nu}p^\rho p^\sigma\nonumber\\
\!\!\!&&\!\!\!+\,\frac{1}{2\Delta_1}\frac{yf^2}{g^2}\,x^2(x^2-2x+1-y^2)p^\mu p^\nu p^\rho p^\sigma\Bigg\}.
\end{eqnarray}

\section{Renormalization and Finite Terms}\label{ren1}
\subsection{$\beta$- and $\gamma$-Functions}\label{bet}
Before extracting the divergences of this theory, we must
define an appropriate set of renormalization conditions. These conditions
will also be necessary to compute finite corrections, which will be done in
section {\ref{fin1}}. We will use the standard $\phi^{\phantom{n}\!\!\!4}$ theory renormalization conditions:
\begin{center}
\includegraphics[scale=1]{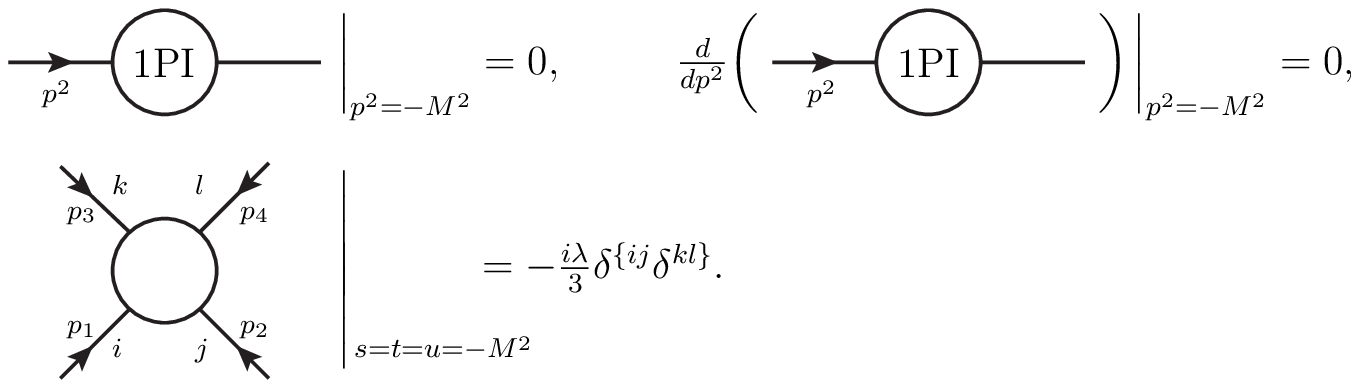}
\end{center}
Using the results found in section \ref{sec-2} and the renormalization conditions
defined above, we find (neglecting their finite contributions)
\begin{eqnarray}\label{ren1.1}
\delta_\lambda\!\!\!&=&\!\!\!\frac{\lambda^2(N+8)}{6(4\pi)^2}\big[\Gamma(\epsilon/2)-\ln M^2\big]+\mathcal{O}(\lambda^3)
=-2\delta_{\lambda_1}=8\delta_{\lambda_2}=4\delta_{\lambda_3},\nonumber\\
\delta_Z\!\!\!&=&\!\!\!-\frac{\lambda^2(N+2)}{36(4\pi)^4}\big[\Gamma({\epsilon})-\ln M^2\big]+\mathcal{O}(\lambda^3)
=-\delta_{Z_1}=-\delta_{Z_2}=2\delta_{Z_3}=2\delta_{Z_4}=\delta_{Z_5},\nonumber\\
\delta_{Z_6}\!\!\!&=&\!\!\!0.
\end{eqnarray}
The $\mathcal{O}(\lambda^3)$ corrections in the vertex and the propagator counterterms come
from two- and three-loop order corrections respectively. A two-loop approximation will be
used in the $\beta$-function calculation. (However, we shall not discuss the mass renormalization.)
These results confirm renormalizability of the theory. There are no other divergences at one-loop
order, the relations among the counterterms are consistent with the $O(N)$ symmetry, and all nonlocal
divergences are canceled at two-loop order.

Using the relations (\ref{ren1.1}) and the Callan-Symanzik
equation (CSE)
\begin{eqnarray}\label{ren1.2}
\left[M\frac{\partial}{\partial M}+\beta(\lambda)\frac{\partial}{\partial\lambda}+n\gamma(\lambda)\right]
G^{(n)}(p_1,p_2,\dots,p_n)\bigg\vert_{-M^2}=0,
\end{eqnarray}
we find the $\beta$- and $\gamma$-functions
\begin{eqnarray}\label{ren1.3}
\beta(\lambda)\!\!\!&=&\!\!\!\frac{\lambda^2(N+8)}{3(4\pi)^2}\,\Pi
-\frac{\lambda^3(3N+14)}{3(4\pi)^4}\,\Pi\,^2+\mathcal{O}(\lambda^4)\\
\gamma(\lambda)\!\!\!&=&\!\!\!\frac{\lambda^2(N+2)}{36(4\pi)^4}\,\Pi^{\,2}+\mathcal{O}(\lambda^3),
\end{eqnarray}
where $\Pi=\left(1-\frac{1}{2}K_{\mu\nu}\eta^{\mu\nu}
+\frac{1}{8}K_{\mu\nu}K_{\alpha\beta}\eta^{\{\mu\nu}\eta^{\alpha\beta\}}\right)+\mathcal{O}(K^3)$.

Although the function $\Pi$ was computed up to second order in $K$,
it can be generalized to any order. 
For this purpose, we must compute the infinite contribution to $V^{(n)}$, which 
represents the one-loop correction to the scalar vertex when $n$ insertions of $K$ are introduced. 
A detailed explanation can be found in the appendix. The result is 
\begin{eqnarray}\label{ren1.4}
V^{(n)}_{\,\infty}=\frac{(-1)^{n+1}}{2(4\pi)^2}\left(\frac{1}{2}\right)^n\frac{1}{n!}\,\Gamma(\epsilon/2)\,
K_{\mu_1\nu_1}K_{\mu_2\nu_2}\cdots K_{\mu_n\nu_n}\eta^{\{\mu_1\nu_1}\eta^{\mu_2\nu_2}\cdots \eta^{\mu_n\nu_n\!\}}.
\end{eqnarray}
with the symmetrized sum $K_{\mu_1\nu_1}K_{\mu_2\nu_2}\cdots K_{\mu_n\nu_n}\eta^{\{\mu_1\nu_1}\eta^{\mu_2\nu_2}\dots \eta^{\mu_n\nu_n\}}
\equiv \,\hat{\phantom{n}\!\!\!\!\!s}^{\phantom{n}\!\!\!n}[K_{\mu\nu}\eta^{\mu\nu}]$
as defined by eq.~(\ref{app.4}) in the appendix.
Once the contributions from the different channels are included, the $K$-independent term 
$V^{(0)}_{\,\infty}=-\frac{\Gamma(\epsilon/2)}{2(4\pi)^2}$ reproduces the $\mathcal O(\lambda^2)$ Lorentz-invariant term
of eq.~(\ref{ren1.3}). Therefore, the remaining $n$-dependent factor in eq.~(\ref{ren1.4})
is the $\mathcal{O}(K^n)$ contribution to $\Pi$. Adding all the contributions at all orders (as shown in the appendix), we find
\begin{eqnarray}\label{ren1.5}
\Pi\!\!\!&=&\!\!\!
\frac{1}{\sqrt{\textrm{det}(\,\mathds{1}+K)}}\,,
\end{eqnarray}
which was also found in the low energy regime in~\cite{ref-Brett-1}.
Using this result, we see that the $\beta$- and $\gamma$-functions rescale as
\begin{eqnarray}\label{ren1.6}
\beta(\lambda)=\sum_{n=2}^\infty\lambda^n\beta_{n}^{(0)}\big[\,\textrm{det}(\,\mathds{1}+K)\big]^{-(n-1)/2},\hspace{0.5cm}
\gamma(\lambda)=\sum_{n=2}^\infty\lambda^n\gamma_{n}^{(0)}\big[\,\textrm{det}(\,\mathds{1}+K)\big]^{-n/2},
\end{eqnarray}
where $\beta_n^{(0)}$ and $\gamma_n^{(0)}$ are the $\mathcal{O}(\lambda^n)$ results for
the Lorentz-invariant theory.

This result is to be expected. To understand why this rescaling is possible, we consider the action for the theory
\begin{eqnarray}\label{ren1.7}
S=\int d^4x\left\{\frac{1}{2}\big[\,\eta^{\mu\nu}+K^{\mu\nu}\big]\partial_\mu\phi_i\,\partial_\nu\phi_i
-\frac{1}{2}\mu^2\phi^2_i-\frac{\lambda}{4!}(\phi_i^2)^2\right\}.
\end{eqnarray}
Under the change in coordinates, $x'^\mu=x^\mu-\frac{1}{2}K^\mu_\nu x^\nu$, 
the partial derivatives transform as $\partial\phantom{n}\!\!\!'_{\!\mu}=\partial_\mu+\frac{1}{2}K^{\nu}_{\,\mu}\partial_\nu$.
Under this transformation, the $K$-dependence in eq.~(\ref{ren1.7}) 
can be eliminated, although the transformation has a nontrivial Jacobian.
To first order in $K$, $J=\left|\textrm{det}\left(\frac{\partial x^\mu}{\partial x'^\nu}\right)\right|=
\left|\textrm{det}\big(\delta^\mu_\nu+\frac{1}{2}K^\mu_\nu\big)\right|$.
Including the correct transformation and Jacobian to all orders, eq.~(\ref{ren1.7}) becomes
\begin{eqnarray}\label{ren1.8}
S\!\!\!&=&\!\!\!\int d^4x'\sqrt{\textrm{det}(\,\mathds{1}+K)}
\left[\,\frac{1}{2}(\partial\phantom{n}\!\!\!'_{\!\mu}\phi_{\phantom{n}\!\!\!i})^{2}
-\frac{1}{2}\,\mu^2\phi^{\phantom{n}\!\!\!2}_{\phantom{y}\!\!\!i}
-\frac{\lambda}{4!}(\phi_{\phantom{y}\!\!\!i}^{\phantom{n}\!\!\!2})^2\right]\nonumber\\
\!\!\!&=&\!\!\!\int d^4x'\left[\frac{1}{2}(\partial\phantom{n}\!\!\!'_{\!\mu}\phi\phantom{n}\!\!\!'_{\phantom{n}\!\!\!\!i}\phantom{n}\!\!\!)^2
-\frac{1}{2}\,\mu^2\phi\phantom{n}\!\!\!'^{\phantom{y}\!\!\!2}_{\phantom{y}\!\!\!\!i}
-\frac{\lambda\phantom{n}\!\!\!'}{4!}(\phi\phantom{n}\!\!\!'^{\phantom{y}\!\!\!2}_{\phantom{n}\!\!\!\!i})^2
\right].
\end{eqnarray}
where
\begin{eqnarray}\label{ren1.9}
\phi\phantom{n}\!\!\!'_{\phantom{n}\!\!\!\!i}=\big[\textrm{det}(\,\mathds{1}+K)\big]^{1/4}\phi_{\phantom{n}\!\!\!i}\,,\,\,\,\,\,\,
\lambda\phantom{n}\!\!\!'=\big[\textrm{det}(\,\mathds{1}+K)\big]^{-1/2}\lambda\,.
\end{eqnarray}
The relations (\ref{ren1.6}) can then be deduced after using only the redefinitions (\ref{ren1.9}).
Note that this means that when there are only scalar fields, the theory with $K$ is
equivalent to the Lorentz-invariant $\phi^{\phantom{n}\!\!\!4}$ theory, albeit in different coordinates.

\subsection{Finite Corrections}\label{fin1}

In this section, we will present the finite contributions to the two- and four-point functions,
using the renormalization conditions given in section \ref{bet}.
Using these conditions, and neglecting terms of the order
$\mathcal O(K^2)$, we find
\begin{eqnarray}\label{fin1.1}
Z_2^{(1)}(p\cdot p\,,\,\mu^2,M^2)\!\!\!&=&\!\!\!-\frac{(N+2)\,\Pi^{\,2}}{2(4\pi)^4}\Bigg[\,\frac{p^2+M^2}{2}
-\!\int_0^1\!\!dx\!\int_0^{1-x}\!\!\!\!dy\,\frac{\Delta_1(p^2)}{g^2}
\ln\left(\frac{\Delta_1(p^2)}{\Delta_1(-M^2)}\right)\nonumber\\
\!\!\!&+&\!\!\!3K_{\mu\nu}p^\mu p^\nu
\int_0^1\!\!dx\int_0^{1-x}\!\!\!dy\,\frac{yf^{\,2}}{g^4}
\bigg\{\ln\left(\frac{\Delta_1(p^2)}{\Delta_1(-M^2)}\right)-\frac{yf}{g}\frac{p^2+M^2}{\Delta_1(-M^2)}\bigg\}\Bigg]\nonumber\\
\!\!\!&=&\!\!\!\Pi^{\,2} Z_2^{(0)}\big(p\circ p\,,\,\mu^2,M^2-K_{\mu\nu}p^\mu p^\nu\big).
\end{eqnarray}
In~(\ref{fin1.1}), $p\cdot p$ and $p\circ p$ indicate different inner products of $p$ with itself.
For the first case, $p_{\phantom{n}\!\!\!i}\cdot p_j=\eta_{\mu\nu}p_i^\mu p_j^\nu$, while $p_i\circ p_j=(\eta_{\mu\nu}+K_{\mu\nu})p_i^\mu p_j^\nu$ 
for the second. Since the renormalization conditions are defined at $p^2=-M^2$, it is natural that 
$M^2$ rescales as $M^2\rightarrow M^2-K_{\mu\nu}p^\mu p^\nu$.

The lowest-order finite correction to the vertex is
\begin{eqnarray}\label{fin1.2}
V^{(1)}(p)\!\!\!&=&\!\!\!\frac{\Pi}{2(4\pi)^2}
\Bigg\{\int_0^1\!dx\ln\left[\frac{\Delta(p)}{\Delta(-M^2)}\right]
-2K_{\mu\nu}p^\mu p^\nu\int_0^1\!dx\frac{(1-x)^2x^3(p^2+M^2)}{\Delta(p^2)\Delta(-M^2)}\Bigg\}\nonumber\\
\!\!\!&=&\!\!\!\Pi\,V^{(0)}\big(p\circ p\,,\,\mu^2,M^2-K_{\mu\nu}p^\mu p^\nu\big).
\end{eqnarray}

Eqs.~(\ref{fin1.1}) and (\ref{fin1.2}) can be generalized to
any order in $K$, using the relationship to the conventional theory discussed in section \ref{bet}.
Adding the contributions from all loop orders,
the quantum corrections to the two- and four-point functions for the 
Lorentz-violating theory can be written as
\begin{eqnarray}\label{fin1.3}
G^{(2)ij}_K\!\!\!&=&\!\!\!\prod_{l=1}^2\frac{i\,\delta^{ij}}{p\circ p-\mu^2}
\sum_{q=1}^{\infty}Z_q\big(\phantom{n}\!\!\!\Pi\lambda\,,\{p\circ p\},\mu^2,M^2-K_{\mu\nu}p^\mu p^\nu\big)\\\label{fin1.3a}
G^{(4)ijkl}_K\!\!\!&=&\!\!\!\prod_{n=1}^4\frac{i\,\Pi^{\,-1}}{p_n\circ p_n-\mu^2}
\sum_{q=1}^{\infty}V_q^{ijkl}\big(\phantom{n}\!\!\!\Pi\lambda\,,\{p_i\circ p_j\},\mu^2,M^2-K_{\mu\nu}p^\mu p^\nu\big),
\end{eqnarray}
where $Z_q$ and $V_q^{ijkl}$ are the $q$-loop contributions (with no external legs)
to the propagator and vertex of
the Lorentz-invariant theory, respectively. Eqs.~(\ref{fin1.3}) and~(\ref{fin1.3a}) can easily be generalized to any $n$-point function,
which is in agreement with the transformations~(\ref{ren1.9}).

\section{Lorentz Violation in Yukawa Interactions}\label{yuk}
\subsection{SME Lagrangian}\label{sec-yuk-0}
Having studied the pure scalar sector in detail, we will now turn our attention to theories that include
fermions and Yukawa interactions, with dimension 2, 3 and 4 operators
within the SME framework. We start by writing the general Lagrange density
\begin{eqnarray}\label{yuk.1}
\mathcal L_Y\!\!\!&=&\!\!\!
\frac{1}{2}\partial^\mu\phi\partial_\mu\phi
+\frac{1}{2}K^{\mu\nu}\partial_\mu\phi\partial_\nu\phi+u^\beta\phi\partial_\beta\phi
+\phi^{2}v^\beta\partial_\beta\phi
-V(\phi)\nonumber\\
\!\!\!&&\!\!\!+\,\bar\psi\!\left(i\Gamma^\mu\partial_\mu-M\right)\psi
-\phi\bar\psi G\psi,
\end{eqnarray}
where
\begin{eqnarray}\label{yuk.2}
\Gamma^\nu\!\!\!&=&\!\!\!\gamma^\nu+\Gamma_1^\nu=\gamma^\nu+c^{\mu\nu}\gamma_\mu+d^{\mu\nu}\gamma_5\gamma_\mu+e^{\nu}
+i\gamma_5f^\nu+\frac{1}{2}g^{\lambda\mu\nu}\sigma_{\lambda\mu}\\
M\!\!\!&=&\!\!\!m+i\gamma_5m'+M_1=m+i\gamma_5m'+a^\mu\gamma_\mu+b^\mu\gamma_5\gamma_\mu+\frac{1}{2}H^{\mu\nu}\sigma_{\mu\nu}\label{yuk.2a}\\
G\!\!\!&=&\!\!\!g+i\gamma_5g'+G_1=g+i\gamma_5g'+I^\mu\gamma_\mu+J^\mu\gamma_5\gamma_\mu+\frac{1}{2}L^{\mu\nu}\sigma_{\mu\nu}\label{yuk.2b}.
\end{eqnarray}
In QED, the LV at the fermion-boson vertex is entirely determined by the coefficients in the free fermion
sector~\cite{ref-kost2}. This is a consequence of gauge invariance in QED, but the same requirement
does not apply here. Instead, there may be distinct $I^\mu$, $J^\mu$ and $L^{\mu\nu}$ coefficients.
The Lorentz violation in the pure scalar sector
is described by the symmetric matrix $K_{\mu\nu}$, as before, although we shall specialize to the particular case $N=1$.
The terms $u^\beta\phi\partial_\beta\phi=\frac{1}{2}\partial_{\beta}(u^\beta\phi^2)$ 
and $\phi^{\phantom{n}\!\!\!2}v^\beta\partial_\beta\phi=\frac{1}{3}\partial_{\beta}(v^\beta\phi^3)$ in eq.~(\ref{yuk.1})
can again be dropped for constant $u^\beta$ and $v^\beta$. 

A list of the discrete symmetry properties of the operators shown in eqs.~(\ref{yuk.1}--\ref{yuk.2b}) is shown
in table \ref{tab-yuk-1}. The mixing of operators under quantum corrections is constrained by these symmetries.
However, the situation is more complex than in QED, because the Lorentz-invariant operators parametrized by $m'$
and $g'$ are odd under P and T. Consequently, operators with different P and T symmetries may mix, through
multiplication by $m'$ and $g'$.

The potential $V(\phi)$ describes the interaction among the scalar
fields; it will be given as usual by
\begin{eqnarray}\label{yuk.3}
V(\phi)=\frac{1}{2}\mu^2\phi^2+\frac{1}{4!}\lambda\phi^4,
\end{eqnarray}
with $\lambda>0$.
\begin{table}
\begin{center}
{\renewcommand{\arraystretch}{1.6}
\renewcommand{\tabcolsep}{0.2cm}
\begin{tabular}{|c|c|c|c|c|c|c|c|}
\hline
\textrm{Operator} & \textrm{C}  & \textrm{P}  & \textrm{T} &
\textrm{CP}       & \textrm{CT} & \textrm{PT} & \textrm{CPT}\\
\hline
$g,\,m,\,c_{00},\,c_{ij},\,K_{\mu\nu}$ 
& $+$ & $+$ & $+$ & $+$ & $+$ & $+$ & $+$\\
\hline
$b_j,\,J_j,\,g_{i0k},\,g_{ij0}$
& $+$ & $+$ & $-$ & $+$ & $-$ & $-$ & $-$\\
\hline
$b_0,\,J_0,\,g_{i00},\,g_{ijk}$
& $+$ & $-$ & $+$ & $-$ & $+$ & $-$ & $-$\\
\hline
$g',\,m',\,c_{0j},\,c_{j0}$
& $+$ & $-$ & $-$ & $-$ & $-$ & $+$ & $+$\\
\hline
$a_0,\,I_0,\,e_0,\,f_j$
& $-$ & $+$ & $+$ & $-$ & $-$ & $+$ & $-$\\
\hline
$H_{ij},\,L_{ij},\,d_{j0},\,d_{j0}$
& $-$ & $+$ & $-$ & $-$ & $+$ & $-$ & $+$\\
\hline
$H_{0j},\,L_{0j},\,d_{00},\,d_{ij}$
& $-$ & $-$ & $+$ & $+$ & $-$ & $-$ & $+$\\
\hline
$a_{j},\,I_j,\,e_j,\,f_{0}$
& $-$ & $-$ & $-$ & $+$ & $+$ & $+$ & $-$\\
\hline
\end{tabular}}
\caption{Discrete symmetry properties.}
\label{tab-yuk-1}
\end{center}
\end{table}

\begin{figure}
\centering
\includegraphics[scale=1]{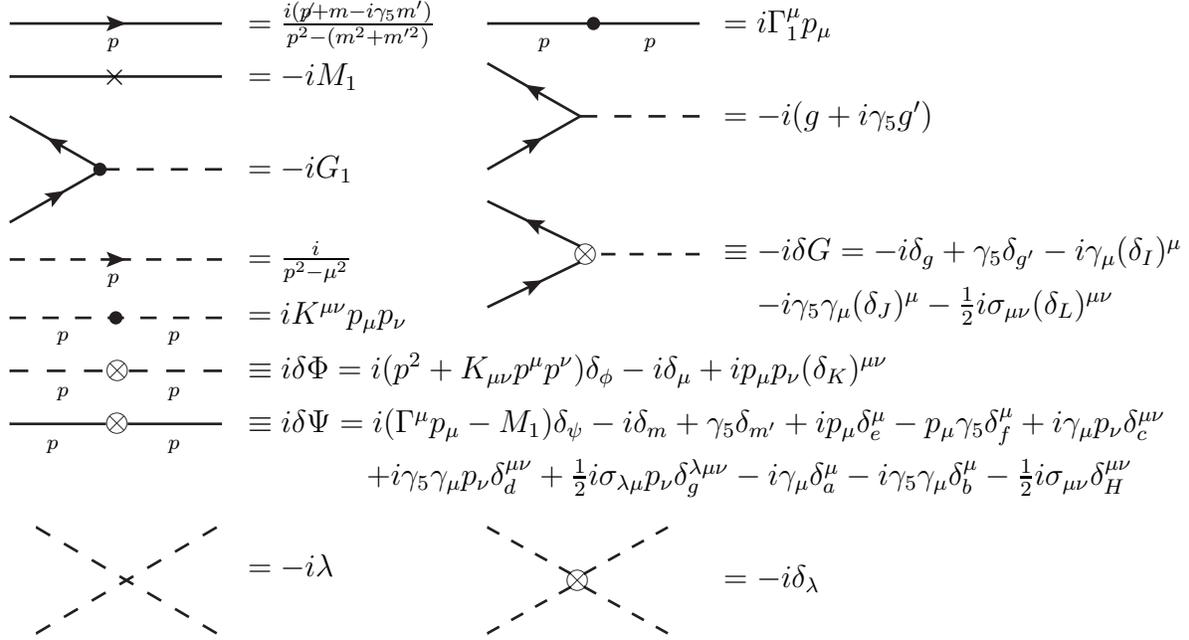}
\caption{Feynman rules for the Yukawa theory with LV.}
\label{fig-f.rules1}
\end{figure}

As usual, quantum corrections modify the propagation and interactions of the different fields.
The calculation of these corrections as well as the divergences that determine the behavior of the 
$\beta$-functions will be studied in the next sections. The Feynman rules for
this theory are summarized in fig. \ref{fig-f.rules1}.

\subsection{The Scalar Propagator}\label{yuk-1}
\begin{figure}
\centering
\includegraphics[scale=1]{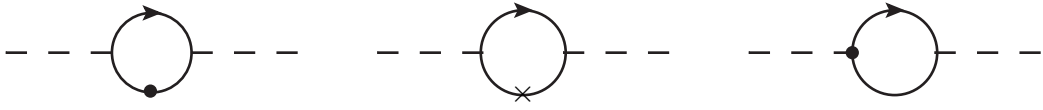}
\caption{Lowest order diagrams with LV that contribute to $Z(p)$. There are another
three diagrams that contribute---when the Lorentz-violating insertion is introduced
in the opposite internal line or vertex. However, they give the same results.}
\label{fig-yuk-1a}
\end{figure}

At one-loop order, we will consider all the possible corrections to
the scalar propagator. They can be divided in two groups, with each group related
to a specific vertex---scalar or Yukawa. The total one-loop contribution to the 
scalar propagator---including external leg corrections---will be written as 
\begin{eqnarray}\label{yuk-1.a}
\frac{i}{p^2+K_{\mu\nu}p^\mu p^\nu-\mu^2}iZ(p)\frac{i}{p^2+K_{\mu\nu}p^\mu p^\nu-\mu^2}
\end{eqnarray}
where $Z(p)$ is the sum of all one-particle-irreducible (1PI) insertions into
the scalar propagator. 
The contribution given by the pure scalar sector, $Z^K(p)$, was previously computed, 
so we will only show the results for the Yukawa vertices.
We will focus our attention on the divergences, which determine
the behavior under the RG.
The correction coming from the Yukawa sector will be written
as $Z(p)=Z^0(p)+\sum_x Z^x(p)$, where $Z^0(p)$ is the Lorentz-invariant
contribution, given below by eq.~(\ref{yuk-1.b}), and $Z^x(p)$ includes the
Lorentz-violating contribution associated with the coefficient $x$.
The diagrams that contribute to the various $Z^x(p)$ are illustrated in fig.~\ref{fig-yuk-1a}. 
The Lorentz-invariant contribution is
\begin{eqnarray}\label{yuk-1.b}
i Z^0\!\!&=&\!\!\!\frac{4i}{(4\pi)^2}\int_0^1\!\!dx\Bigg\{\frac{\Gamma(\epsilon/2)}{\Delta_2^{\epsilon/2}}
\Big[\,3(g^2+g'^{\,2})x(1-x)p^2-(3g^2+g'^{\,2})m^2-(\,g^2+3g'^{\,2})m'^{\,2}\Big]\Bigg\}\nonumber\\
\!\!\!&-&\!\!\!\frac{4i(g^2+g'^{\,2})}{(4\pi)^2}\Big(m^2+m'^{\,2}-\frac{1}{6}p^2\Big)\nonumber\\
\!\!\!&+&\!\!\!\frac{ig'^{\,2}}{(4\pi)^2\epsilon}\,\textrm{Tr}\Big[\gamma_5\{\gamma_\alpha,\gamma_5\}\gamma_\beta\Big]
\Big[(m^2+m'^{\,2})\eta^{\alpha\beta}-\frac{1}{6}p^2\eta^{\alpha\beta}-\frac{1}{3}p^\alpha p^\beta\Big].
\end{eqnarray}
where $\Delta_2=m^2+m'^{\,2}-x(1-x)p^2$. 

The last two terms in eq.~(\ref{yuk-1.b}) are finite corrections. For a theory
in $d=4$ dimensions, $\{\gamma_5,\gamma_\nu\}=0$; however, the fact that $d=4-\epsilon$, introduces an $\mathcal O(\epsilon)$
correction in $\{\gamma_5,\gamma_\nu\}$. This is an analogue to the axial vector anomaly in a gauge theory.
The infinite contribution to eq.~(\ref{yuk-1.b}) is
\begin{eqnarray}\label{yuk-1.c}
iZ_\infty^0=4i\eta\left[\,\frac{1}{2}(g^2+g'^{\,2})p^2-(3g^2+g'^{\,2})m^2
-4gg'mm'-(g^2+3g'^{\,2})m'^{\,2}\right],
\end{eqnarray}
where $\eta\equiv\frac{\Gamma(\epsilon/2)}{(4\pi)^2}$.
The Lorentz-violating infinite contributions are:
\begin{eqnarray}\label{yuk-1.d}
iZ^{c}_\infty\!\!\!&=&\!\!\!4i\eta \,c^{\mu\nu}\,\bigg[-\frac{1}{2}(g^2+g'^{\,2})p^2\eta_{\mu\nu}
+(g^2+g'^{\,2})p_\mu p_\nu+m^2(3g^2+g'^{\,2})\eta_{\mu\nu}\nonumber\\
\!\!\!&&\!\!\!+\,4gg'mm'\eta_{\mu\nu}+m'^{\,2}(g^2+3g'^{\,2})\eta_{\mu\nu}\bigg]\\
iZ^{e}_\infty\!\!\!&=&\!\!\!-8i\eta\, g(gm+g'm')e^\mu p_\mu\\
iZ^{f}_\infty\!\!\!&=&\!\!\!-8i\eta\, g'(gm+g'm')f^\mu p_\mu\\
iZ^{a}_\infty\!\!\!&=&\!\!\!8i\eta(g^2+g'^{\,2})a^\mu p_\mu\\
iZ^d_\infty\!\!\!&=&\!\!\!iZ^g_\infty=iZ^b_\infty=iZ^H_\infty=iZ^I_\infty=iZ^J_\infty=iZ^L_\infty=0.
\end{eqnarray} 
The terms $Z^e_\infty, Z^f_\infty$ and $Z^a_\infty$ are unimportant total derivatives that can be dropped.
In a more general theory that included spacetime-dependent Lorentz-violating coefficients,
terms of the form $p_\mu Q^\mu(\{y_i\})$ could no longer be discarded, because they would
not represent total derivatives. In this case, they would generate quantum corrections to
the term $u^\alpha\phi\partial_\alpha\phi$.
 
\subsection{Fermion Propagator}\label{yuk-2}

At the lowest order, the fermion propagator is only corrected 
by the emission and reabsorption of a virtual scalar field.
Its one-loop Lorentz-invariant contribution is
\begin{eqnarray}\label{yuk-2.a}
-i\Sigma\!\!\!&=&\!\!\!\frac{i}{(4\pi)^2}\int_0^1dx\frac{\Gamma(\epsilon/2)}{\Delta_3^{\epsilon/2}}
\big[(g+i\gamma_5g')^2(m-i\gamma_5 m')+(g^2+g'^{\,2})x\!\!\not\!p\big]\nonumber\\
\!\!\!&&\!\!\!-\,\frac{g'(g+i\gamma_5g')}{(4\pi)^2\epsilon}\{\not\!p\,,\gamma_5\},
\end{eqnarray}
with $\Delta_3=x\mu^2+(1-x)(m^2+m'^{,2})-x(1-x)p^2$.
The infinite part of eq.~(\ref{yuk-2.a}) is
\begin{eqnarray}\label{yuw-2.b}
-i\Sigma_\infty=i\eta\,
\left\{\frac{1}{2}(g^2+g'^{\,2})\!\!\not\!p+(g^2-g'^{\,2})m+2gg'm'-i\gamma_5\Big[(g^2-g'^{\,2})m'-2gg'm\,\Big]\right\}.
\end{eqnarray}
The infinite Lorentz-violating contributions to the fermion self-energy can be divided into the
ones coming from the insertions on the vertices, and on the scalar and fermion propagators.
The three types give respectively
\begin{figure}
\centering
\includegraphics[scale=1]{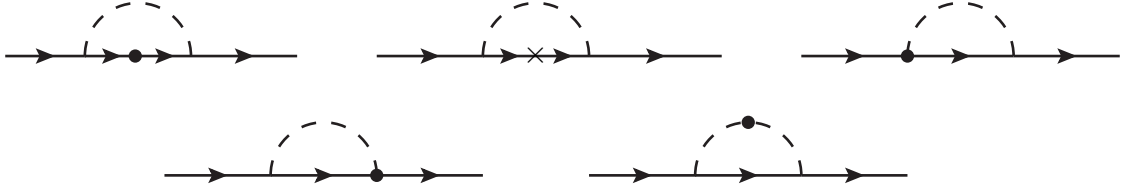}
\caption{Lorentz-violating one-loop contributions to the fermion self-energy.}
\label{fig-yuk-1b}
\end{figure}
\begin{eqnarray}\label{yuk-2.c1}
-i\Sigma^{I}_\infty\!\!\!&=&\!\!\!i\eta I^{\mu}
\Big[g\,p_\mu+i\gamma_5g'p_\mu+2(gm+g'm')\gamma_\mu\Big]\label{yuk-2.c1a}\\
-i\Sigma^{J}_\infty\!\!\!&=&\!\!\!i\eta J^{\mu}
\Big[-\frac{1}{2}g\,\varepsilon_{\mu\nu\alpha\beta}\,\sigma^{\alpha\beta}p^\nu+g'\sigma_{\mu\nu}p^\nu
+2(gm+g'm')\gamma_5\gamma_\mu\Big]\label{yuk-2.c1b}\\
-i\Sigma^{L}_\infty\!\!\!&=&\!\!\!i\eta\,\frac{1}{2}L^{\mu\nu}
\Big[-g\gamma_5\,\varepsilon_{\mu\nu\alpha\beta}\,\gamma^\beta p^\alpha-2g'\gamma_5p_\mu\gamma_\nu
+2(gm+g'm')\sigma_{\mu\nu}\nonumber\\
\!\!\!&&\!\!\!+(g'm-gm')\varepsilon_{\mu\nu\alpha\beta}\,\sigma^{\alpha\beta}\Big]\\\label{yuk-2.c1c}
-i\Sigma^{K}_\infty\!\!\!&=&\!\!\!-i\eta K^{\mu\nu}
\Bigg\{\frac{1}{6}(g^2+g'^{\,2})(\eta_{\mu\nu}\!\!\not\!p-p_\mu\gamma_\nu)
+\frac{1}{4}\Big[(g^2-g'^{\,2})m+2gg'm'\Big]\eta_{\mu\nu}\nonumber\\
\!\!\!&&\!\!\!-\frac{i}{4}\gamma_5\Big[(g^2-g'^{\,2})m'-2gg'm\Big]\eta_{\mu\nu}\Bigg\}\label{yuk-2.c2}
\end{eqnarray}
\begin{eqnarray}\label{yuk-2.c3}
-i\Sigma^{c}_\infty\!\!\!&=&\!\!\!-i\eta\,c^{\mu\nu}
\Bigg\{\frac{1}{6}(g^2+g'^{\,2})(\eta_{\mu\nu}\!\!\not\!p+p_\mu\gamma_\nu-2\gamma_\mu p_\nu)
+\frac{1}{2}\Big[(g^2-g'^{\,2})m+2gg'm'\Big]\eta_{\mu\nu}\nonumber\\
\!\!\!&&\!\!\!-\frac{i}{2}\gamma_5\Big[(g^2-g'^{\,2})m'-2gg'm\Big]\eta_{\mu\nu}\Bigg\}\label{yuk-2.c3a}\\
-i\Sigma^{d}_\infty\!\!\!&=&\!\!\!i\eta\,d^{\mu\nu}
\Bigg\{\frac{1}{6}\gamma_5(g^2+g'^{\,2})(\eta_{\mu\nu}\!\!\not\!p+p_\mu\gamma_\nu-2\gamma_\mu p_\nu)\nonumber\\
\!\!\!&&\!\!\!+\frac{1}{4}\Big[(g^2-g'^{\,2})m+2gg'm'\Big]\varepsilon_{\mu\nu\alpha\beta}\,\sigma^{\alpha\beta}
+\frac{1}{2}\Big[(g^2-g'^{\,2})m'-2gg'm\Big]\sigma_{\mu\nu}\Bigg\}\label{yuk-2.c3b}\\
-i\Sigma^{e}_\infty\!\!\!&=&\!\!\!-\frac{1}{2}i\eta e^\mu\,
\Big[(g^2-g'^{\,2})p_\mu+2i\gamma_5gg'p_\mu+(g^2+g'^{\,2})m\gamma_\mu\Big],\label{yuk-2.c3c}\\
-i\Sigma^{f}_\infty\!\!\!&=&\!\!\!\frac{1}{2}i\eta\,f^{\mu}\Big[i\gamma_5(g^2-g'^{\,2})p_\mu
-2gg'p_\mu-(g^2+g'^{\,2})m'\gamma_\mu\Big]\label{yuk-2.c3d}\\
-i\Sigma^{g}_\infty\!\!\!&=&\!\!\!-i\eta\,g^{\lambda\mu\nu}
\Bigg\{\frac{1}{12}(g^2-g'^{\,2})\Big[\sigma_{\lambda\mu}p_\nu+2\eta_{\lambda\nu}\sigma_{\mu\beta}p^\beta-2\sigma_{\lambda\nu}p_\mu\Big]
-\frac{1}{4}(g^2+g'^{\,2})m\gamma_5\varepsilon_{\lambda\mu\nu\beta}\gamma^\beta\nonumber\\
\!\!\!&&\!\!\!+\frac{1}{12}gg'\Big[\varepsilon_{\lambda\mu\alpha\beta}\sigma^{\alpha\beta}p_\nu
+2\eta_{\lambda\nu}\varepsilon_{\mu\beta\alpha\rho}\,\sigma^{\alpha\rho}p^\beta
-2\varepsilon_{\lambda\nu\alpha\rho}\,\sigma^{\alpha\rho}p_\mu\Big]\nonumber\\
\!\!\!&&\!\!\!-\frac{1}{2}(g^2+g'^{\,2})m'\gamma_5\eta_{\lambda\nu}\gamma_\mu\Bigg\}\\\label{yuk-2.c3e}
-i\Sigma^{a}_\infty\!\!\!&=&\!\!\!-\frac{1}{2}i\eta(g^2+g'^{\,2})a^\mu \gamma_\mu\label{yuk-2.c4a}\\
-i\Sigma^{b}_\infty\!\!\!&=&\!\!\!\frac{1}{2}i\eta(g^2+g'^{\,2})b^\mu \gamma_5\gamma_\mu\label{yuk-2.c4b}\\
-i\Sigma^H_\infty\!\!\!&=&\!\!\!0\label{yuk-2.c4c}.
\end{eqnarray}

\subsection{Yukawa Vertex Corrections}\label{yuk-3}

\begin{figure}
\centering
\includegraphics[scale=1]{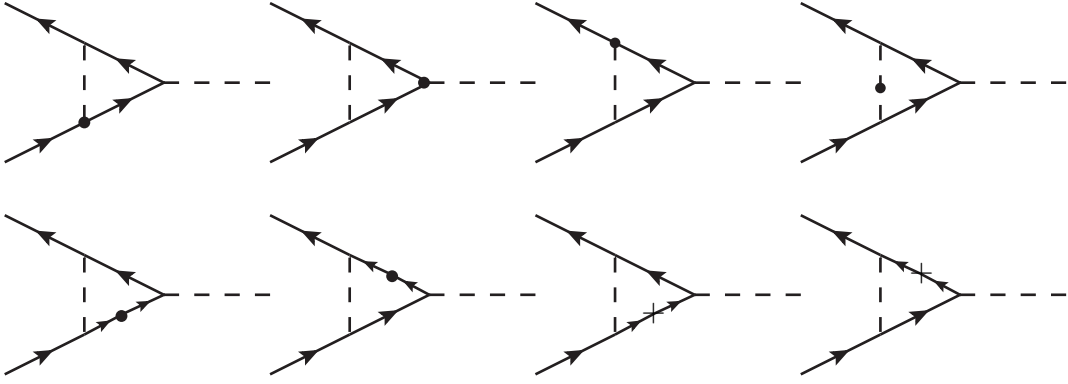}
\caption{Lorentz-violating corrections to the Yukawa vertex.}
\label{fig-yuk-3.a}
\end{figure}
We now consider the quantum correction to the Yukawa vertex, beginning
with the usual vertex correction with no
Lorentz-violating terms. Denoting as $p$, $p'$, and $q=p'-p$ the 
momenta of the incoming fermion, outgoing fermion and incoming boson 
respectively, we find
\begin{eqnarray}\label{yuk-3.a}
i\mathcal G\!\!\!&=&\!\!\!2i\eta(g+i\gamma_5g')(g^2+g'^{\,2})
\!\int_0^1\! dx\int_0^{1-x}\!\!dy\frac{1}{\Delta_4^{\epsilon/2}}
-\frac{i(g^2+g'^{\,2})}{2(4\pi)^2}(g+i\gamma_5g')\\
\!\!\!&&\!\!\!-\,\frac{i(g+i\gamma_5g')}{(4\pi)^2}
\int_0^1\!\!dx\int_0^{1-x}\!\!dy
\frac{1}{\Delta_4}\,\Bigg\{(\not l+m-i\gamma_5m')
\Big[(g^2+g'^{\,2})(\not l\,+\!\not\!q)\nonumber\\
\!\!\!&&\!\!\!+\,(m-i\gamma_5m')(g+i\gamma_5g')^2\Big]\Bigg\}\nonumber
-\frac{g'(g^2+g'^{\,2})}{2(4\pi)^2\epsilon}\gamma^\alpha\,\{\gamma_\alpha,\gamma_5\},
\end{eqnarray}
where we have defined $l\equiv p(1-x)-yp'$ and $\Delta_4\equiv
\mu^2(1-x-y)+(x+y)(m^2+m'^{\,2})-x(1-x)p^2+2xyp'\cdot p-y(1-y)p'^{\,2}$.
In terms of $\xi\equiv\frac{\Gamma(\epsilon/2)}{(4\pi)^2}(g^2+g'^{\,2})$, 
the infinite part of eq.~(\ref{yuk-3.a}) is just 
\begin{eqnarray}\label{yuk-3.b}
i\mathcal G_\infty=i\xi(g+i\gamma_5g').
\end{eqnarray}
Now we want to solve the eight diagrams shown in fig.~\ref{fig-yuk-3.a} that include the LV. 
They produce the infinite results
\begin{eqnarray}\label{yuk-3.c}
i\mathcal G^{I}_\infty\!\!\!&=&\!\!\!\frac{3}{2}i\xi I^\mu\gamma_\mu\label{yuk-3.ca}\\
i\mathcal G^{J}_\infty\!\!\!&=&\!\!\!\frac{5}{2}i\xi J^\mu\gamma_5\gamma_\mu\label{yuk-3.cb}\\
i\mathcal G^{L}_\infty\!\!\!&=&\!\!\!i\xi L^{\mu\nu}\sigma_{\mu\nu}\\\label{yuk-3.cc}
i\mathcal G^{K}_\infty\!\!\!&=&\!\!\!-\frac{1}{4}i\xi(g+i\gamma_5g')K^{\mu\nu}\eta_{\mu\nu}\label{yuk-3.cd}
\end{eqnarray}
\begin{eqnarray}\label{yuk-3.ce}
i\mathcal G^{c}_\infty\!\!\!&=&\!\!\!-\frac{1}{2}i\xi(g+i\gamma_5g')c^{\mu\nu}\eta_{\mu\nu}\label{yuk-3.cf}\\
i\mathcal G^{d}_\infty\!\!\!&=&\!\!\!-\frac{1}{4}i\xi
d^{\mu\nu}\Big(-g\,\varepsilon_{\mu\nu\alpha\beta}\,\sigma^{\alpha\beta}+2g'\sigma_{\mu\nu}\Big)\label{yuk-3.cg}\\
i\mathcal G^{e}_\infty\!\!\!&=&\!\!\!-\frac{1}{2}i\xi ge^{\mu}\gamma_{\mu}\label{yuk-3.ch}\\
i\mathcal G^{f}_\infty\!\!\!&=&\!\!\!-\frac{1}{2}i\xi g'\,f^{\mu}\gamma_{\mu}\label{yuk-3.ci}\\
i\mathcal G^{g}_\infty\!\!\!&=&\!\!\!\frac{1}{4}i\xi g^{\lambda\mu\nu}
\Big(g\gamma_5\,\varepsilon_{\lambda\mu\nu\beta}\gamma^\beta+2g'\gamma_5\eta_{\lambda\nu}\gamma_\mu\Big)\\\label{yuk-3.cj}
i\mathcal G^{a}_\infty\!\!\!&=&\!\!\!i\mathcal G^{b}_\infty=i\mathcal G^{H}_\infty=0\label{yuk-3.ck}.
\end{eqnarray}

\subsection{Scalar Vertex}\label{yuk-4}
\begin{figure}
\centering
\includegraphics[scale=1]{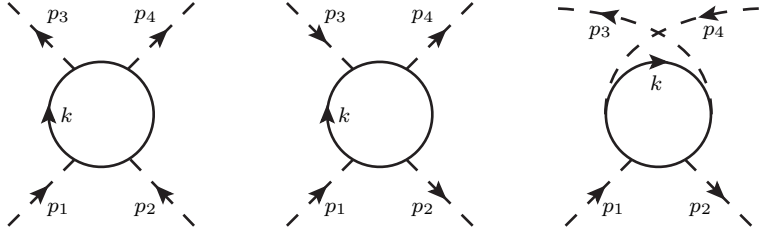}
\caption{Three of the six Lorentz-invariant diagrams involving a fermion loop and four external scalar legs. 
The other three are obtained by flipping the direction of the fermion current.}
\label{fig-yuk-4a}
\end{figure}

In addition to the corrections coming from the pure scalar sector
discussed in section \ref{sec-2.1}, there are extra corrections to the
$\phi^4$ vertex, coming from diagrams with fermion loops. At the lowest order, there are a total
of six different diagrams that contribute; three of them are shown in fig.~\ref{fig-yuk-4a}.
The complete expression for the diagrams shown in fig.~\ref{fig-yuk-4a} is extremely long, even
in the Lorentz-invariant case. Nevertheless, obtaining the divergences is straightforward, since the diagrams
can have at most logarithmic divergences. Taking into account 
the six different contributions, the infinite part for the Lorentz-invariant case is
\begin{eqnarray}\label{yuk-4.a}
iV_\infty=-24i\eta(g^2+g'^{\,2})^2.
\end{eqnarray}
The infinite Lorentz-violating corrections only receive contributions from $c_{\mu\nu}$, and
the result is
\begin{eqnarray}\label{yuk-4.b}
iV^{c}_\infty=24i\eta(g^2+g'^{\,2})^2\,c^{\mu\nu}\eta_{\mu\nu}.
\end{eqnarray}

\section{Renormalization}\label{ren2}
\subsection{Renormalization Conditions}\label{condit}
To renormalize the theory, an appropriate set of renormalization conditions
must be introduced. For the scalar propagator and vertex, we will use the renormalization
conditions shown in section \ref{bet}. For the fermion propagator and Yuwawa vertex, we will use the
conditions
\begin{center}
\includegraphics[scale=1]{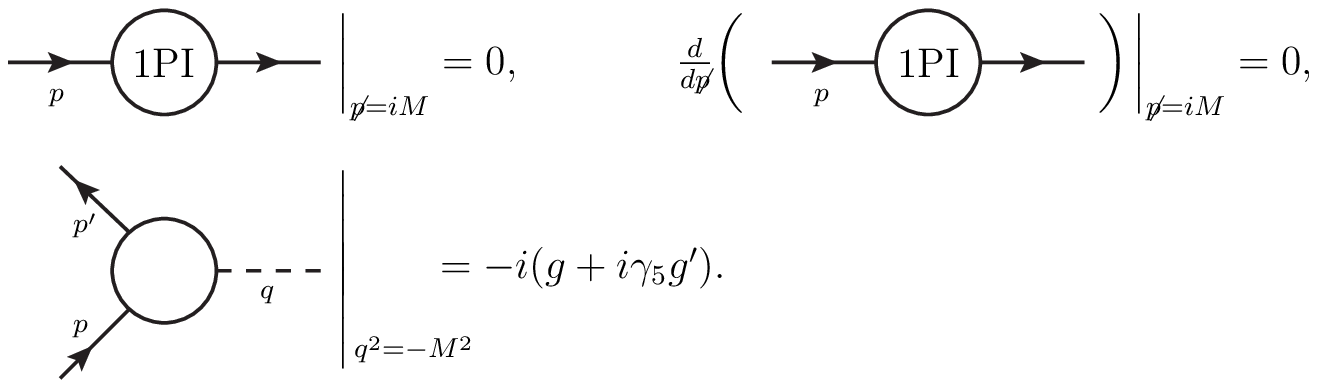}
\end{center}

Let $\Xi$ be a general one-loop correction. In the massless limit ($m^2,\,\mu^2\ll M^2$),
its contribution evaluated at $M$ can be written as
\begin{eqnarray}\label{ren2.b}
\Xi\equiv \Xi_\infty\big[\Gamma(\epsilon/2)-\ln M^2+\textrm{finite}\big],
\end{eqnarray}
where $\Xi_\infty$ is its divergent coefficient, which determines the counterterms.
A list of these counterterm coefficients can be found in tables \ref{tab-1}, \ref{tab-2}
and \ref{tab-3}.
\begin{table}
\begin{center}
{\renewcommand{\arraystretch}{1.5}
\renewcommand{\tabcolsep}{0.1cm}
\begin{tabular}{|c|l|}
\hline
Operator & \hspace{3.1cm}Scalar Field Counterterm\\
\hline
$ p^2 $   & $\delta_\phi=-2\eta(g^2+g'^{\,2})$\\
\hline
$\mathds{1}$ & $\delta_{\mu^2}=-4\eta[\,m^2(3g^2+g'^{\,2})+4gg'mm'+m'^{\,2}(g^2+3g'^{\,2})](1-c^{\,\nu}_{\,\,\,\nu})$\\
&$\hspace{1.06cm}-\lambda\zeta(1-\frac{1}{2}K^{\nu}_{\,\,\nu})$ \\ 
\hline
$p_\mu p_\nu$ &  $\delta_{K}^{\mu\nu}=-2\eta(g^2+g'^{\,2})(\,c^{\mu\nu}+c^{\nu\mu}-c^{\,\alpha}_{\,\,\,\alpha}\eta^{\mu\nu}-K^{\mu\nu})$\\
\hline
\end{tabular}}
\end{center}
\caption{One-loop counterterms for the scalar sector, in terms of $\eta=\frac{\Gamma(\epsilon/2)}{(4\pi)^2}$ and 
$\zeta=\frac{\Gamma(-1+\epsilon/2)(\mu^2)^{1-\epsilon/2}}{2(4\pi)^{2}}$.} 
\label{tab-1}
\end{table}
\begin{table}
\begin{center}
{\renewcommand{\arraystretch}{1.5}
\renewcommand{\tabcolsep}{0.1cm}
\begin{tabular}{|c|l|}
\hline
Operator & \hspace{3.6cm}Fermion Field Counterterm\\
\hline
$\not\!p$   & $\delta_\psi=-\frac{1}{2}\eta(\phantom{n}\!\!\!g^2+g'^{\,2})
$\\
\hline
$\mathds{1}$ & $\delta_m=\eta \big[(\,g^2-g'^{\,2})m+2gg'm'\big](1-\frac{1}{2}c^{\mu}_{\,\,\mu}-\frac{1}{4}K^{\mu}_{\,\,\mu})$\\
\hline
$\gamma_5$ & $\delta_{m'}=\eta \big[\,2gg'm-(g^2-g'^{\,2})m'\,\big](1-\frac{1}{2}c^{\,\alpha}_{\,\,\,\alpha}-\frac{1}{4}K^\alpha_{\,\,\alpha})$\\
\hline
$p_\mu$ & $\delta_{e}^\mu=-\eta[\,g I^\mu -g^2e^\mu-gg'f^\mu]$ \\
\hline
$\gamma_5 p_\mu$ & $\delta_{f}^\mu=-\eta[\,g'I^\mu-g'^{\,2}f^\mu-gg' e^\mu]$ \\
\hline
$\gamma_\mu p_\nu$ & $\delta_{c}^{\mu\nu}=\frac{1}{6}\eta(g^2+g'^{\,2})[\,c^{\mu\nu}+c^{\nu\mu}-K^{\mu\nu}
+\eta^{\mu\nu}(c^\alpha_{\,\,\alpha}+K^\alpha_{\,\,\alpha})]$ \\
\hline
$\gamma_5 \gamma_\mu p_\nu$ & $\delta_{d}^{\mu\nu}=-\eta[\,g'L^{\mu\nu}-\frac{1}{6}(\,g^2+g'^{\,2})(5d^{\mu\nu}
-d^{\nu\mu}-\eta^{\mu\nu}d^{\alpha}_{\,\,\alpha})
-\frac{1}{2}g\,\varepsilon^{\alpha\beta\nu\mu}L_{\alpha\beta}]$ \\
\hline
$\sigma_{\lambda\mu}p_\nu$ & $\delta_{g}^{\lambda\mu\nu}=\eta[\,\frac{1}{3}(2g^2+g'^{\,2})g^{\lambda\mu\nu}
+\frac{1}{3}(g^2-g'^{\,2})(g^{\beta\lambda}_{\,\,\,\,\,\,\beta}\,\eta^{\mu\nu}-g^{\lambda\nu\mu})
+g\,\varepsilon^{\alpha\nu\lambda\mu}J_\alpha$ \\
& $\hspace{1.3cm}-2g'J^{\lambda}\eta^{\mu\nu}+\frac{1}{6}gg'(\,g_{\alpha\beta}^{\,\,\,\,\,\,\,\,\nu}\,\varepsilon^{\alpha\beta\lambda\mu}
+2g^\alpha_{\,\,\,\beta\alpha}\,\varepsilon^{\beta\nu\lambda\mu}+2g^\nu_{\,\,\,\sigma\rho}\varepsilon^{\sigma\rho\lambda\mu})]$ \\
\hline
$\gamma_\mu$ & $\delta_{a}^\mu=\eta[\,2(gm+g'm')I^\mu-\frac{1}{4}(g^2+g'^{\,2})(2me^\mu+2m'f^\mu)]$ \\
\hline
$\gamma_5\gamma_\mu$ & $\delta_{b}^\mu=\eta[\,2(gm+g'm')J^\mu
+\frac{1}{4}(g^2+g'^{\,2})(4b^\mu+2m'g^{\lambda\mu}_{\,\,\,\,\,\,\,\lambda}+m \,g_{\lambda\beta\nu}\,\varepsilon^{\lambda\beta\nu\mu})]$ \\
\hline
$\sigma_{\mu\nu}$ & $\delta_{H}^{\mu\nu}=\eta[\,\frac{1}{2}(g^2+g'^{\,2})H^{\mu\nu}
+\frac{1}{2}[(g^2-g'^{\,2})m+2gg'm'\,]\,\varepsilon^{\alpha\beta\mu\nu}d_{\alpha\beta}$\\
& $\hspace{0cm}+[(g^2-g'^{\,2})m'-2gg'm]d^{\mu\nu}
+2(gm+g'm')L^{\mu\nu}+(g'm-gm')\,\varepsilon^{\alpha\beta\mu\nu}L_{\alpha\beta}]$ \\
\hline
\end{tabular}}
\end{center}
\caption{One-loop counterterms for the free fermion sector.} 
\label{tab-2}
\end{table}
\begin{table}
\begin{center}
{\renewcommand{\arraystretch}{1.5}
\renewcommand{\tabcolsep}{0.1cm}
\begin{tabular}{|c|l|}
\hline
Operator & \hspace{1.5cm}Yukawa Vertex Counterterm\\
\hline
$\mathds{1}$   & $\delta_g=\xi g(1-\frac{1}{2}c^{\,\mu}_{\,\,\,\mu}-\frac{1}{4}K^\mu_{\,\,\mu})$\\
\hline

$\gamma_5$ &  $\delta_{g'}=\xi g'(1-\frac{1}{2}c^{\,\mu}_{\,\,\,\mu}-\frac{1}{4}K^\mu_{\,\,\mu})$\\
\hline
$\gamma_\mu$ & $\delta_{I}^\mu=\frac{1}{2}\xi(\,3I^\mu-ge^\mu-g'f^\mu)$ \\
\hline
$\gamma_5\gamma_\mu$ &  $\delta_{J}^{\mu}=\frac{1}{4}\xi
(\,10J^\mu+g\,\varepsilon^{\alpha\beta\nu\mu}\,g_{\alpha\beta\nu}+2g'g^{\lambda\mu}_{\,\,\,\,\,\,\lambda}\,)$\\
\hline
$\sigma_{\mu\nu}$ & $\delta_{L}^{\mu\nu}=\frac{1}{2}\xi
(\,4L^{\mu\nu}+g\,\varepsilon^{\alpha\beta\mu\nu}\,d_{\alpha\beta}-2g'd^{\mu\nu}\,)$ \\
\hline
& \hspace{1.5cm}Scalar Vertex Counterterm\\
\hline
$\mathds{1}$ & $\delta_{\lambda}=\frac{3}{2}\eta[\,\lambda^2(1-\frac{1}{2}K^{\mu}_{\,\,\mu})
-16(\,g^2+g'^{\,2})^2(1-c^{\,\mu}_{\,\,\,\mu})]$\\
\hline
\end{tabular}}
\end{center}
\caption{One-loop counterterms for the scalar and Yukawa interaction vertices.} 
\label{tab-3}
\end{table}

\subsection{$\beta$-Functions}\label{beta}

After finding the full set of counterterms, we can compute the $\beta$-function
associated with each operator. These determine how
the coupling constants evolve with the momentum scale. For this purpose we will need
the CSE
\begin{eqnarray}\label{beta.1}
\left[M\frac{\partial}{\partial M}+\sum_{x_i}\beta_{x_i}\frac{\partial}{\partial x_i}
+n\gamma_\phi+m\gamma_\psi\right]G^{(n,m)}(\{p_{\,i}\},\{x_i\},M)\Bigg\vert_{-M^2}=0,
\end{eqnarray}
where $\beta_{x_i}$ is the $\beta$-function associated with a parameter $x_i$ (for instance $c^{\mu\nu}$, $b^\mu$, etc.).
The functions $\gamma_\phi$ and $\gamma_\psi$ are the usual $\gamma$-functions for the scalar and fermion
fields. The renormalization scale $M$ comes into play in the counterterms,
which cancel the divergences of the theory.

In the massless limit, we can describe the $M$-dependence of the counterterms $\delta_{x_i}$ by
shifting $\eta$ as $\eta\rightarrow\eta_M=\frac{1}{(4\pi)^2}[\,\Gamma(\epsilon/2)-\ln M^2+\textrm{finite}\,]$, 
where the finite contribution does not depend on $M$. Therefore,
the counterterms are shifted according to $\delta_{x_i}(\eta)\rightarrow \delta_{x_i}(\eta_M)$, and this implies that
\begin{eqnarray}\label{beta.2}
M\frac{\partial}{\partial M}\Big[\delta_{x_i}(\eta_M)\Big]=-\frac{2}{\Gamma(\epsilon/2)}\delta_{x_i}(\eta)
\equiv -2\bar\delta_{x_i}.
\end{eqnarray}
where $\bar\delta_{x_i}$ is just the finite factor that multiplies the divergent factor $\Gamma(\epsilon/2)$, i.e.,
$\Xi_\infty$ in eq.~(\ref{ren2.b}).

We will explain in detail how the CSE can be applied to the scalar two-point
function, $G^{(2)}(p^2,\{x_i\},M)$. When the CSE is applied to the other
correlation functions, the remaining $\beta$-functions---which cannot be determined from $G^{(2)} $ alone---can 
also be computed.

The scalar two-point function can be written as the free correlation function plus the one--loop corrections
and the counterterms (including the external legs). It is given by
\begin{eqnarray}\label{beta.3}
G^{(2)}(\phantom{i}\!p^2,\{x_i\},M)=D_F^1(p)
+D_F^1(p)\big[iZ(p)\big]D^1_F(p)+D_F^1(p)\big[i\delta\Phi(M)\big]D_F^1(p),
\end{eqnarray}
where $D_F^1(p)$ is the full free Lorentz-violating scalar field propagator, given by
\begin{eqnarray}\label{beta.4}
D_F^1(p)=\frac{i}{p^2-\mu^2}+\frac{i}{p^2-\mu^2}iK^{\mu\nu}p_\mu p_\nu\frac{i}{p^2-\mu^2}+\mathcal{O}(K^2)
\equiv D_F^0(p)\big[1-\Omega(K)\big],
\end{eqnarray}
with $D_F^0(p)$ the Lorentz-invariant propagator.
$iZ(p)$ is the total one loop correction (including all the Lorentz-violating contributions), and $i\delta\Phi$ is the
scalar field counterterm. We will treat the masses as small perturbations; this
means that $\frac{\mu^2}{M^2}\ll1$ and $\frac{m^2}{M^2}\ll1$. The renormalization conditions
previously stated in section \ref{bet} and \ref{condit} imply that
\begin{eqnarray}\label{beta.5}
iZ(p\,,\{x_i\})\Bigg\vert_{p^2=-M^2}+i\delta\Phi(M)\!\!\!&=&\!\!\!0\\\label{beta.5a}
\frac{\partial}{\partial x_i}\Big[iZ(p,\{x_i\})+i\delta\Phi(M,\{x_i\})\Big]\Bigg\vert_{p^2=-M^2}\!\!\!&=&\!\!\!0.
\end{eqnarray}
Now we can apply the CSE to eq.~(\ref{beta.3}). The $M\frac{\partial}{\partial M}$ operator in the CSE only acts
on the counterterm; therefore
\begin{eqnarray}\label{beta.6}
M\frac{\partial}{\partial M}\,G^{(2)}\Bigg\vert_{-M^2}=
\big[D_F^1(-M^2)\big]^2M\frac{\partial}{\partial M}\,i\delta\Phi(M)\Bigg\vert_{p^2=-M^2}.
\end{eqnarray}
The $\sum_i\beta_{x_i}\frac{\partial}{\partial x_i}$ will only act on $D_F^1(p)$, because the
contribution from the one-loop diagrams and the counterterm cancel each other after evaluation at $p^2=-M^2$. 
Since the free Lorentz-violating propagator $D_F^1(p)$ only depends on $\mu$ and $K^{\mu\nu}$, we find
\begin{eqnarray}\label{beta.7}
\sum_{i}\beta_{x_i}\frac{\partial}{\partial x_i}G^{(2)}\Bigg\vert_{-M^2}\!\!\!\!\!\!\!
\!\!\!&=&\!\!\!
-i\big[D_F^0(-M^2)\big]^2\big[1-\Omega(K,-M^2)\big]\beta_{\mu^2}\nonumber\\
\!\!\!&&\!\!\!+i\big[D_F^0(-M^2)\big]^2p_\alpha p_\beta (\beta_K)^{\alpha\beta}.
\end{eqnarray}
The last term of the CSE gives
\begin{eqnarray}\label{beta.8}
2\gamma_\phi G^{(2)}\Bigg\vert_{-M^2}=2\gamma_\phi D_F^1(-M^2).
\end{eqnarray}
Using these three contributions---eqs.~(\ref{beta.6}--\ref{beta.8})---we find after some simplification
\begin{eqnarray}\label{beta.9}
\Big[1-\Omega(-M^2)\Big]M\frac{\partial}{\partial M}\delta\Phi\Bigg\vert_{-M^2}-\beta_{\mu^2}
+p_\mu p_\nu(\beta_K)^{\mu\nu}+2(M^2+\mu^2)\gamma_\phi+\mathcal{O}(K^2)=0,
\end{eqnarray}
which becomes
\begin{eqnarray}\label{beta.10}
\!\!\!&-&\!\!\!M^2M\frac{\partial}{\partial M}\delta_\phi-M\frac{\partial}{\partial M}\delta_{\mu^2}
+p_\alpha p_\beta\frac{\partial}{\partial M}(\delta_{K})^{\alpha\beta}-\beta_{\mu^2}\nonumber\\
\!\!\!&+&\!\!\!p_\alpha p_\beta(\beta_K)^{\alpha\beta}+2(M^2+\mu^2)\gamma_\phi
+\mathcal{O}\Big(K\frac{\mu^2}{M^2}\Big)=0.
\end{eqnarray}
Comparing powers of $M^2$ and momentum, we see
\begin{eqnarray}\label{beta.11}
\gamma_\phi\!\!\!&=&\!\!\!\frac{1}{2}M\frac{\partial}{\partial M}\Big[\delta_{\phi}(\eta_M)\Big]=-\bar\delta_\phi,\nonumber\\
\beta_{\mu^2}\!\!\!&=&\!\!\!-M\frac{\partial}{\partial M}\Big[\delta_{\mu^2}(\eta_M)-\mu^2\delta_\phi(\eta_M)\Big]
=2\Big[\bar\delta_{\mu^2}(\zeta=0)-\mu^2\bar\delta_\phi\Big],\nonumber\\
(\beta_{K})^{\mu\nu}\!\!\!&=&\!\!\!-M\frac{\partial}{\partial M}\Big[\delta_{K}^{\mu\nu}(\eta_M)\Big]=2\bar\delta_{K}^{\mu\nu}.
\end{eqnarray}
As mentioned above, the remaining counterterms can be found applying the CSE to the other correlation functions.
For the fermion two-point function the process is similar to the one just explained; however, the fermion self-energy
contains many more operators. The vertices demand more work, since we must include corrections on the external 
legs, using the expressions already found for the boson and fermion self-energies.
The $\beta$- and $\gamma$-functions
of the fermion field operators are.
\begin{eqnarray}\label{beta.12}
\gamma_\psi\!\!\!&=&\!\!\!-\bar\delta_\psi,\\
\beta_m\!\!\!&=&\!\!\!2(\bar\delta_m-m\bar\delta_\psi)\\
\beta_{m'}\!\!\!&=&\!\!\!2\bar\delta_{m'}\\\label{beta.12a}
(\beta_{x_i})^{\mu_1,\dots,\mu_n}\!\!\!&=&\!\!\!2(\bar\delta_{x_i})^{\mu_1,\dots,\mu_n}.
\end{eqnarray}
On the other hand, the $\beta$-functions for the operators associated with the Yukawa and scalar vertices are
\begin{eqnarray}\label{d3}
\beta_g\!\!\!&=&\!\!\!2\bar\delta_g-g\bar\delta_\phi-2g\bar\delta_{\psi}\\
\beta_{g'}\!\!\!&=&\!\!\!2\bar\delta_{g'}-g'\bar\delta_\phi-2g'\bar\delta_{\psi}\\
(\beta_{I})^\mu\!\!\!&=&\!\!\!2\bar\delta_{I}^\mu-I^\mu\bar\delta_\phi-2I^\mu\bar\delta_{\psi}\\
(\beta_{J})^\mu\!\!\!&=&\!\!\!2\bar\delta_{\!J}^\mu-J^\mu\bar\delta_\phi-2J^\mu\bar\delta_{\psi}, \\
(\beta_{L})^{\mu\nu}\!\!\!&=&\!\!\!2\bar\delta_{L}^{\mu\nu}-L^{\mu\nu}\bar\delta_\phi-2L^{\mu\nu}\bar\delta_{\psi}\\\label{d3.a}
\beta_{\lambda}\!\!\!&=&\!\!\!2\bar\delta_\lambda-4\lambda\bar\delta_\phi.
\end{eqnarray}

\subsection{Running Coupling Constants}\label{run}
Finally, we will describe the momentum dependence of the different operators whose
$\beta$-functions were found in section \ref{beta}. Let $x_i^{\mu_1\dots\mu_n}(\tilde p\,;\{x_j\})$ be 
a momentum-dependent
operator at $\tilde p=p/M$; $x_i^{\mu_1\dots\mu_n}$ its value given at $M$ ($\tilde p=1$); and 
$(\beta_{x_i})^{\mu_1\dots\mu_n}$ its $\beta$-function.
Then, $x_i^{\mu_1\dots\mu_n}(\tilde p\,;\{x_j\})$ satisfies the differential equation 
\begin{eqnarray}\label{eq-run.1}
\!\!\!&&\!\!\!\tilde p\frac{d}{d\tilde p}x_i^{\mu_1\dots\mu_n}(\tilde p\,;\{x_j\})=(\beta_{x_i})^{\mu_1\dots\mu_n},\nonumber\\
\!\!\!&&\!\!\!\textrm{with boundary condition:}\,\,\,
x_i^{\mu_1\dots\mu_n}(1\,;\{x_j\})=x_i^{\mu_1\dots\mu_n}.\hspace{1.0cm}
\end{eqnarray}
Solving eq.~(\ref{eq-run.1}) is difficult in general because of the mixing of different operators through their
$\beta$-functions, which produces a system of coupled nonlinear differential equations.
Since most of the $\beta$-functions depend on $g$ and $g'$, we should solve for them first.
Using eq.~(\ref{eq-run.1}), we find
\begin{eqnarray}\label{eq-run.2}
\tilde p\,\frac{dg}{d\tilde p}\!\!\!&=&\!\!\!\frac{g(g^2+g'^{\,2})}{(4\pi)^2}
\left(5-\frac{1}{2}c^{\alpha}_{\,\,\,\alpha}-\frac{1}{4}K^{\alpha}_{\,\,\,\alpha}\right)\\\label{eq-run.2a}
\tilde p\,\frac{dg'}{d\tilde p}\!\!\!&=&\!\!\!\frac{g'(g^2+g'^{\,2})}{(4\pi)^2}
\left(5-\frac{1}{2}c^{\alpha}_{\,\,\,\alpha}-\frac{1}{4}K^{\alpha}_{\,\,\,\alpha}\right).
\end{eqnarray}
Since $|K^\alpha_{\,\,\,\alpha}|,|c^\alpha_{\,\,\,\alpha}|\ll1$, they can be neglected in eqs.~(\ref{eq-run.2}) 
and~(\ref{eq-run.2a}). Defining the function
\begin{eqnarray}\label{eq-run.3}
F(\tilde p)\,\equiv\,1-\frac{5}{(4\pi)^2}(g^2+g'^{\,2})\ln\tilde p^{\,2},
\end{eqnarray}
we find
\begin{eqnarray}\label{eq-run.4}
g(\tilde p)=g\big[F(\tilde p)\big]^{-1/2}=\frac{g}{g'}\,g'(\tilde p).
\end{eqnarray}
Solving for the renormalization flow of the other operators is more difficult, 
and in most of the cases, analytical solutions have not
been found. 
One of the operators that does offer a partial analytical solution is $d^{\mu\nu}$. Using eq.~(\ref{eq-run.1}),
we find that $d^{\mu\nu}$ satisfies the differential equation
\begin{eqnarray}\label{eq-run.5}
\tilde p\,\frac{d}{d\tilde p}\,d^{\mu\nu}=
-\frac{2}{(4\pi)^2}\Big[\!\!\!\!\!&&\!\!\!\!g'L^{\mu\nu}-\frac{1}{6}(\,g^2+g'^{\,2})(5 d^{\mu\nu}
- d^{\nu\mu}-\eta^{\mu\nu} d^{\alpha}_{\,\,\alpha})
+\frac{1}{2}\,g\,\varepsilon^{\alpha\beta\mu\nu}L_{\alpha\beta}\Big].
\end{eqnarray}
Tracing over $d^{\mu\nu}$, we find that $\tilde p\frac{d}{d\tilde p}d^{\alpha}_{\,\,\,\alpha}=0$,
which means that $d^\alpha_{\,\,\,\alpha}(\tilde p)=d^{\alpha}_{\,\,\,\alpha}$. Moreover, since
$L^{\mu\nu}$ is antisymmetric, the symmetric part of $d^{\mu\nu}$, $d_S^{\mu\nu},$ satisfies the
differential equation
\begin{eqnarray}\label{eq-run.6}
\tilde p\,\frac{d}{d\tilde p}d^{\mu\nu}_S(\tilde p)=
\frac{(g^2+g'^{\,2})}{3(4\pi)^2}\big[F(\tilde p)\big]^{-1}
\big[4d^{\mu\nu}_S(\tilde p)-\eta^{\mu\nu}d^{\alpha}_{\,\,\,\alpha}\big].
\end{eqnarray}
Eq.~(\ref{eq-run.6}) can be integrated to give
\begin{eqnarray}\label{eq-run.7}
d^{\mu\nu}_S(\tilde p)-\frac{1}{4}\eta^{\mu\nu}d^{\alpha}_{\,\,\,\alpha}=
\left(d^{\mu\nu}_S-\frac{1}{4}\eta^{\mu\nu}d^{\alpha}_{\,\,\,\alpha}\right)\big[F(\tilde p)\big]^{-2/15}.
\end{eqnarray}
The antisymmetric part of $d^{\mu\nu}$ is coupled to $L^{\mu\nu}$, making its analytical 
solution much more difficult to find.
The rest of the operators---including the Dirac and Majorana masses---require numerical solutions, which
will not be shown.

An interesting situation is studying the behavior of a given $\beta$-function if only its 
corresponding Lorentz-violating coefficient is nonvanishing. This simplifies the problem and helps us
to figure out the behavior of these operators under the RG.
When each Lorentz-violating operator is considered alone, the $\beta$-function of some 
operators can decouple and be brought to the form 
$(\beta_{x_i})^{\mu_1\dots\mu_n}=f_{x_i}(\phantom{n}\!\!\!g,g')x_i^{\mu_1\dots\mu_n}(\tilde p)$, which implies that
\begin{eqnarray}\label{eq-run.10}
x_i^{\mu_1\dots\mu_n}(\tilde p)=
x_i^{\mu_1\dots\mu_n}\exp\left[\int\frac{d\tilde p}{\tilde p}f_{x_i}\big[\phantom{n}\!\!\!g(\tilde p),g'(\tilde p)\big]\right].
\end{eqnarray}
Therefore, if the differential equation satisfied by the operator $x_i^{\mu_1\dots\mu_n}(\tilde p)$ can be decoupled, the function 
$f_{x_i}\big[\phantom{n}\!\!\!g(\tilde p),g'(\tilde p)\big]$ determines its behavior under the RG; this set
of functions is shown in table~\ref{tab-ren-a}. 
\begin{table}
\begin{center}
{\renewcommand{\arraystretch}{1.4}	
\renewcommand{\tabcolsep}{0.1cm}
\begin{tabular}{|c|c|c|c|c|c|c|c|c|}
\cline{1-2}\cline{4-5}\cline{7-8}
Coefficient & $f_{x_i}\big(\bar\alpha_g,\bar\alpha_{g'})$ & & Coefficient & $f_{x_i}\big(\bar\alpha_g,\bar\alpha_{g'})$ &
& Coefficient & $f_{x_i}\big(\bar\alpha_g,\bar\alpha_{g'})$\\
\cline{1-2}\cline{4-5}\cline{7-8}
$c^{\mu}_{\,\,\mu}$ & $2(\bar\alpha_g+\alpha_{g'})$ & &$e^\mu$ & $2\bar\alpha_{g}$ &
& $J^\mu$ & $8(\bar\alpha_g+\bar\alpha_{g'})$\\
\cline{1-2}\cline{4-5}\cline{7-8}
$c^{\mu\nu}_A$ & $0$ & & $f^\mu$ &  $2\bar\alpha_{g'}$ &
& $L^{\mu\nu}$ & $7(\bar\alpha_g+\bar\alpha_{g'})$\\
\cline{1-2}\cline{4-5}\cline{7-8}
$c^{\mu\nu}_S$ & $\frac{1}{3}(\bar\alpha_{g}+\bar\alpha_{g'})$ & & $a^\mu$ & 0 &
& $K^{\mu}_{\,\,\,\mu}$ &$4(\bar\alpha_{g}+\bar\alpha_{g'})$\\
\cline{1-2}\cline{4-5}\cline{7-8}
$d^{\mu}_{\,\,\mu}$ & $0$ & & $b^\mu$ & $2(\bar\alpha_g+\bar\alpha_{g'})$ & 
& $K^{\mu\nu}_S$ &$4(\bar\alpha_{g}+\bar\alpha_{g'})$\\
\cline{1-2}\cline{4-5}\cline{7-8}
$d^{\mu\nu}_A$ & $2(\bar\alpha_{g}+\bar\alpha_{g'})$ & & $H^{\mu\nu}$ & $(\bar\alpha_g+\bar\alpha_{g'})$ 
\\
\cline{1-2}\cline{4-5}
$d^{\mu\nu}_S$ & $\frac{4}{3}(\bar\alpha_{g}+\bar\alpha_{g'})$ & & $I^\mu$ & $6(\bar\alpha_g+\bar\alpha_{g'})$ 
\\
\cline{1-2}\cline{4-5}
\end{tabular}}
\caption{List of the $f_{x_i}\big(\bar\alpha_g,\bar\alpha_{g'})$ functions
in terms of the couplings $\bar\alpha_g\equiv\frac{g^2}{(4\pi)^2}\big[F(\tilde p)\big]^{-1}$
and $\bar\alpha_{g'}\equiv\frac{g'^{\,2}}{(4\pi)^2}\big[F(\tilde p)\big]^{-1}$. 
For the $c^{\mu\nu}_S$ calculation, the condition $c^{\mu}_{\,\,\mu}=0$ was assumed.}
\label{tab-ren-a}
\end{center}
\end{table}
When the operator $g^{\lambda\mu\nu}$
is considered alone, it satisfies the differential equation
\begin{eqnarray}\label{eq-run.11}
\tilde p\frac{d}{d\tilde p}\,g^{\lambda\mu\nu}\!\!\!&=&\!\!\!\frac{2}{(4\pi)^2}\bigg[\,\frac{1}{3}(2g^2+g'^{\,2})\,g^{\lambda\mu\nu}
+\frac{1}{3}(g^2-g'^{\,2})\left(g^{\beta\lambda}_{\,\,\,\,\,\,\beta}\,\eta^{\mu\nu}-g^{\lambda\nu\mu}\right)\nonumber\\
\!\!\!&&\!\!\!+\,\frac{1}{6}\,gg'\left(\,g_{\alpha\beta}^{\,\,\,\,\,\,\,\,\nu}\,\varepsilon^{\alpha\beta\lambda\mu}
+2g^\alpha_{\,\,\,\beta\alpha}\,\varepsilon^{\beta\nu\lambda\mu}+2g^\nu_{\,\,\,\sigma\rho}\varepsilon^{\sigma\rho\lambda\mu}\right)\bigg],
\end{eqnarray}
which cannot be decoupled, because eq.~(\ref{eq-run.11}) contains operators 
(such us $g_{\alpha\beta}^{\,\,\,\,\,\,\,\nu}\varepsilon^{\alpha\beta\lambda\mu}$ 
and $g^{\beta\lambda}_{\,\,\,\,\,\,\beta}\,\eta^{\mu\nu}$) that have different symmetry properties.

\begin{figure}
\begin{minipage}{0.4\linewidth}
\centering
\includegraphics[scale=0.6]{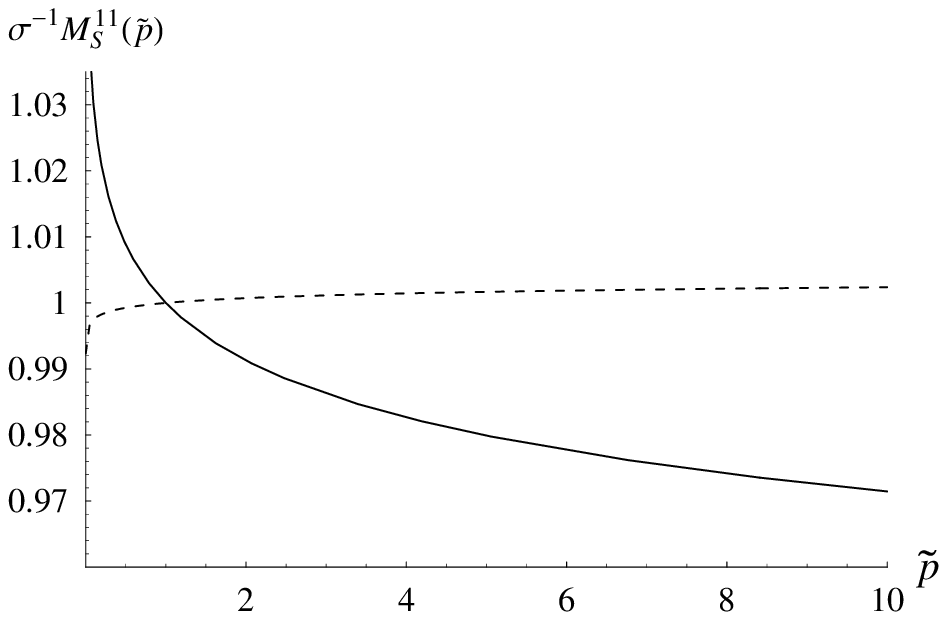}
\end{minipage}
\hspace{0.0cm}
\begin{minipage}{0.4\linewidth}
\centering
\includegraphics[scale=0.6]{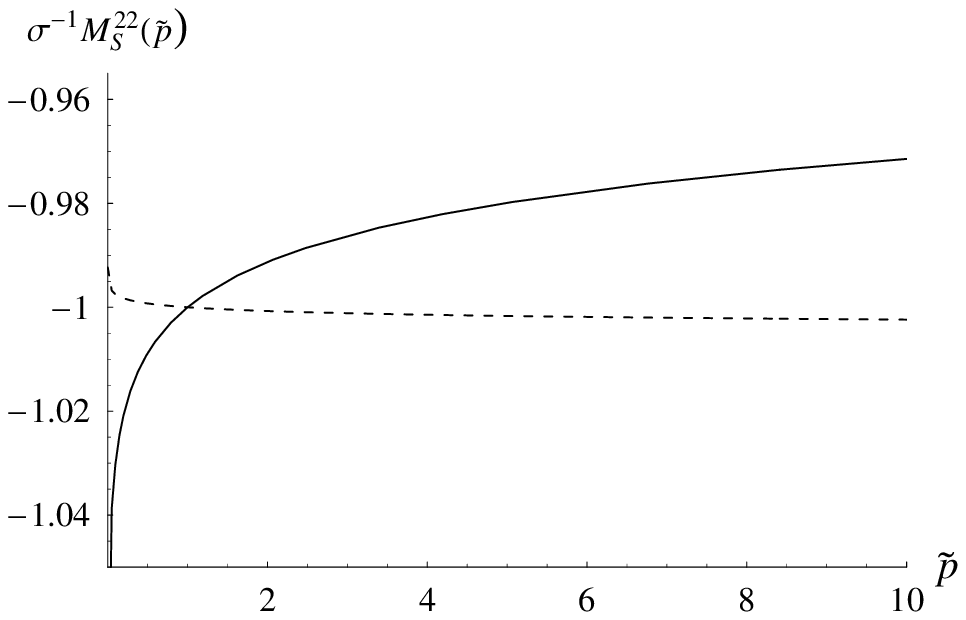}
\end{minipage}
\caption{Plots of $M^{\mu\nu}_S(\tilde p)=K^{\mu\nu}_S(\tilde p)$ (continuous line) and 
$M^{\mu\nu}_S(\tilde p)=c^{\mu\nu}_S(\tilde p)$ (dashed line) versus $\tilde p$,
for $g=g'=0.5$. The graph in the left shows their momentum evolution of the their $\mu=\nu=1$ components, 
while the one in the right describes the behavior of their $\mu=\nu=2$ components.
The initial conditions are $K_S^{\mu\nu}=c^{\mu\nu}_S=\sigma\big(\delta^\mu_1\delta^\nu_1-\delta^\mu_2\delta^\nu_2\big)$,
with $\sigma\ll1$.
Note that the initial conditions imply that both $c$ and $K$ are traceless, $K^\mu_{\,\,\mu}=c^{\mu}_{\,\,\mu}=0$,
and that the remaining components vanish.}
\label{fig-run.a}
\end{figure}

Although every $\beta$-function in table~\ref{tab-ren-a} is positive or zero, the mixing of the operators that arises when
multiple Lorentz-violating operators are considered might change the global sign of the $\beta$-functions
under some conditions. For example, when we consider both $K^{\mu\nu}_S$ and $c^{\mu\nu}_S$ different from
zero (but $K^{\mu}_{\,\,\,\mu}=c^{\mu}_{\,\,\mu}=0$), 
their $\beta$-functions satisfy $(\beta_K)^{\mu\nu}_S=-12(\beta_c)^{\mu\nu}_S$. This condition guarantees that one of
the two operators is asymptotically free, while the other one grows with momentum.This behavior can be 
seen in fig.~{\ref{fig-run.a}}

This generalizes the result found in \cite{ref-anber}, which found that the maximum velocities in different sectors
(equivalent to $c^{00}$ and $K^{00}$) flow toward equality low energies. In fact, this is one aspect of a very general phenomenon.
A traceless symmetric $c$ can, like a $K$, be eliminated (at leading order) by a coordinate redefinition. However, the same
redefinition will work for both the fermion and scalar sectors only if $c^{\mu\nu}+c^{\nu\mu}=K^{\mu\nu}$. If the equality holds,
the LV is unphysical---a mere coordinate artifact. If the equality does not hold, the physically relevant Lorentz-violating parameter is
$c^{\mu\nu}+c^{\nu\mu}-K^{\mu\nu}$. All physical observables, and thus the one-loop $\beta$-function, just depend on that
linear combination. It follows that $c^{\mu\nu}+c^{\nu\mu}=K^{\mu\nu}$ (which represents physical LI) is a fixed
point of the RG.

\section{Conclusions}\label{concl}
First, we studied the behavior of a theory with $N$ scalar fields with mass $m$,
coupled by a $\phi^{\phantom{n}\!\!\!4}$ interaction. The theory was extended to
the Lorentz-violating case by the introduction of the coefficients $K_{\mu\nu}$.
The theory was shown to be perturbatively renormalizable;
eq.~(\ref{ren1.6}) shows how to relate the $\beta$
and $\gamma$ function at any approximation level, once their expressions for the Lorentz-invariant
case are known. To find the finite corrections,
eqs.~(\ref{fin1.3}) and~(\ref{fin1.3a}) can be used; we merely need to find the results in
the Lorentz-invariant theory, which is easier to solve. 

In the pure scalar sector, the effect of the coefficients $K_{\mu\nu}$ is the
one expected. The momenta as well as the renormalization
scale are modified by $p^2\rightarrow p^2+K_{\mu\nu}p^\mu p^\nu$ and
$M^2\rightarrow M^2-K_{\mu\nu }p^\mu p^\nu$. This means that the
Minkowsky metric is effectively modified from $\eta_{\mu\nu}$ to $\eta_{\mu\nu}+K_{\mu\nu}$,
which breaks Lorentz invariance. However, a coordinate transformation can 
make the action Lorentz invariant as long as the coupling constant $\lambda$ and
the fields are rescaled. 
In the trivial case $K_{\mu\nu}=C\eta_{\mu\nu}$, the Lorentz violating and 
Lorentz invariant Lagrangians lead to the same equations of motion under
the rescaling $\phi\rightarrow\sqrt{1+C}\,\phi$, $m^2\rightarrow\frac{m^2}{1+C}$
and $\lambda\rightarrow\frac{\lambda}{(1+C)^2}$. This is consistent with 
eqs.~(\ref{ren1.6}),~(\ref{fin1.3}) and~(\ref{fin1.3a}) by noting that the function 
$\Pi$ becomes $\Pi\rightarrow (1+C)^{-2}$.

With Yukawa coupling with fermions introduced, we
showed how the theory can be renormalized at the one-loop order. 
There are more renormalizable operators in this theory than in Lorentz-violating QED. Moreover, when
both the scalar and pseudoscalar couplings $g$ and $g'$ appear, there is mixing between operators
with different $P$ and $T$ symmetries.

The $\beta$-functions were found applying the CSE to correlation functions. 
The momentum dependence of the Lorentz-violating operators was studied in detail in the
situation where only one form of LV was nonvanishing.
In this case, most of the operators have positive $\beta$-functions, meaning
that their values increase with the energy scale. When various forms of Lorentz-violations are
considered simultaneously, some of the operators mix; their $\beta$-functions are modified and
the overall signs of some $\beta$-functions can change.

\appendix\section*{Appendix: Scalar Vertex to all Orders in $K$}
\renewcommand{\theequation}{A\arabic{equation}}
\setcounter{equation}{0}
Let $V^{(n_1,n_2)}$ represent the one-loop contribution to the scalar vertex when 
$n_1$ and $n_2$ insertions of $K$ are introduced in the first and second internal
lines, respectively. Then
\begin{eqnarray}\label{app.1}
V^{(n_1,n_2)}=\frac{1}{2}\int\frac{d^dq}{(2\pi)^d}
\frac{i(-1)^{n_1}\big[K_{\mu\nu}(q+p)^\mu (q+p)^\nu\big]^{n_1}}{\big[(q+p)^2-\mu^2\big]^{n_1+1}}
\frac{i(-1)^{n_2}\big(K_{\mu\nu}q^\mu q^\nu\big)^{n_2}}{\big(q^2-\mu^2\big)^{n_2+1}}.
\end{eqnarray}
For $n=n_1+n_2$, there are $n+1$ possible ways of distributing the $n$ insertions between
the internal lines. Since $V^{(n)}$ diverges only logarithmically, its infinite
contribution can be found by setting $p=0$; then
\begin{eqnarray}\label{app.2}
V^{(n)}_\infty=\frac{1}{2}(-1)^{n+1}(n+1)\int\frac{d^dq}{(2\pi)^d}
\frac{\big(K_{\mu\nu}q^\mu q^\nu\big)^{n}}{\big(q^2-\mu^2\big)^{n+2}}.
\end{eqnarray}
Performing the Wick rotation $q_0=i\bar q_0$ and $\bar q_i=q_i$, we note that
inside the symmetric integration
\begin{eqnarray}\label{app.3}
\big(K_{\mu\nu}q^\mu q^\nu\big)^{n}=
\frac{(-1)^n (\bar q^2)^n K_{\mu_1\nu_1}K_{\mu_2\nu_2}\cdots K_{\mu_n\nu_n}
\eta^{\{\mu_1\nu_1}\eta^{\mu_2\nu_2}\cdots\eta^{\mu_n\nu_n\!\}}}
{d(d+2)\cdots\big[d+2(n-1)\big]},
\end{eqnarray}
where $\,K_{\mu_1\nu_1}\dots K_{\mu_n\nu_n}\eta^{\{\mu_1\nu_1}\dots\eta^{\mu_n\nu_n\!\}}\,=\,
K_{\{\mu_1\nu_1}\dots K_{\mu_n\nu_n\!\}}\eta^{\mu_1\nu_1}\dots\eta^{\nu_n\nu_n\!}\equiv 
\,\hat{\phantom{n}\!\!\!\!\!s}^{\phantom{n}\!\!\!n}[K_{\mu\nu}\eta^{\mu\nu}]$
is given by
\begin{eqnarray}\label{app.4}
\,\hat{\phantom{n}\!\!\!\!\!s}^{\phantom{n}\!\!\!n}[K_{\mu\nu}\eta^{\mu\nu}]\!\!\!&=&\!\!\! 
K_{\mu_1\nu_1}\cdots K_{\mu_n\nu_n}\eta^{\mu_1\nu_1}\cdots\eta^{\mu_n\nu_n}+
K_{\mu_1\nu_2}\cdots
K_{\mu_n\nu_n}\eta^{\mu_1\nu_1}\cdots\eta^{\mu_n\nu_n}+\dots\nonumber\\
\!\!\!&+&\!\!\!K_{\mu_1\nu_3}\cdots
K_{\mu_n\nu_n}\eta^{\mu_1\nu_1}\cdots\eta^{\mu_n\nu_n}
+\,\textrm{all possible permutations}\nonumber\\
\!\!\!&&\!\!\!\textrm{of the indices on the $K$}
\end{eqnarray}
and represents the $(2n-1)!!$ different ways  of contracting $n$ powers of $K_{\mu\nu}$ and $\eta^{\mu\nu}$.
Noting that $d(d+2)\cdots\big[d+2(n-1)\big]=2^{\,n}\frac{\Gamma(n+d/2)}{\Gamma(d/2)}$, we find that
\begin{eqnarray}\label{app.5}
V^{(n)}_\infty\!\!\!&=&\!\!\!\frac{i(-1)^{n+1}}{2}\frac{(n+1)}{2^{\,n}}
\,\hat{\phantom{n}\!\!\!\!\!s}^{\phantom{n}\!\!\!n}[K_{\mu\nu}\eta^{\mu\nu}]\int\frac{d^d\bar q}{(2\pi)^d}\frac{\Gamma(d/2)}{\Gamma(n+d/2)}
\frac{(\bar q^2)^{n}}{\big(\bar q^2+\mu^2\big)^{n+2}}\,,
\end{eqnarray}
which reproduces eq.~(\ref{ren1.4}).

In order to obtain eq.~(\ref{ren1.5}) we must evaluate
\begin{equation}
\Pi=\sum_{n=0}^{\infty}\frac{(-1)^{n}}{n!}\left(\frac{1}{2}\right)^n\,
\hat{\phantom{n}\!\!\!\!\!s}^{\phantom{n}\!\!\!n}[K_{\mu\nu}\eta^{\mu\nu}].
\end{equation}
$\,\hat{\phantom{n}\!\!\!\!\!s}^{\phantom{n}\!\!\!n}[K_{\mu\nu}\eta^{\mu\nu}]$ 
is the sum over all $(2n-1)$!! distinct contractions of the
$2n$ indices of $K_{\mu_{1}\nu_{1}}\cdots K_{\mu_{n}\mu_{n}}$.
It is simplest to treat $K$ as a matrix, so a cyclic contraction of $n_{i}$ matrices
is ${\rm tr}\,K^{n_{i}}$. Any element in the sum 
$\,\hat{\phantom{n}\!\!\!\!\!s}^{\phantom{n}\!\!\!n}[K_{\mu\nu}\eta^{\mu\nu}]$
will have the form $\prod_{i}({\rm tr}\,K^{n_{i}})^{m_{i}}$, a product of cyclic
contractions, where the cyclic contraction of size $n_{i}$ has multiplicity $m_{i}$.
We must determine combinatorically how many times each 
$\prod_i(\phantom{n}\!\!\!\textrm{tr}\,K^{n_i}\phantom{n}\!\!\!\!)^{m_i}$
appears in $\,\hat{\phantom{n}\!\!\!\!\!s}^{\phantom{n}\!\!\!n}[K_{\mu\nu}\eta^{\mu\nu}]$.
The total number of $K$ matrices appearing is $n=\sum_{i}n_{i}m_{i}$.

There are $\frac{n!}{\prod_{i}(n_{i}!)^{m_{i}}}$ ways of partitioning a set of $n$ elements
into $m_{1}$ distinguishable subsets of size $n_{1}$, $m_{2}$ distinguishable
subsets of size $n_{2}$, etc. However, this overcounts for our purposes, because it
treats the $m_{i}$ sets of size $n_{i}$ as distinguishable. Exchanging all
the $K$ factors in one subset for those in another subset of equal size does not
correspond to a distinct contraction. So the number of ways of choosing the cyclic contractions
is $\frac{n!}{\prod_{i}(n_{i}!)^{m_{i}}m_{i}!}$.

We must also count how many ways there are to contract $n_{i}$ factors of $K$ in a
completely cyclic way. Starting with the first index, there are $2(n_{i}-1)$ ways of
contracting the index with an index on a different $K$. Then, there are $2(n_{i}-2)$
ways of contracting the other index on the $K$ just chosen with an index on a third
$K$. Continuing in this fashion, there are ultimately $2^{n_{i}-1}(n_{i}-1)!$ ways
of forming the contraction. (Alternatively, there are $n_{i}!$ ways of ordering the
$n_{i}$ factors of $K$ and 2 choices of which index to use at each step. However,
this overcounts the number of possibilities by $2n_{i}$, because there are $n_{i}$
cyclic permutations of the $K$ matrices that do not change the overall contraction
structure; switching the choice of index for every $K$ simultaneously also does not produce a new
contraction.)

So we have
\begin{equation}
\,\hat{\phantom{n}\!\!\!\!\!s}^{\phantom{n}\!\!\!n}[K_{\mu\nu}\eta^{\mu\nu}]=\sum_{\lambda}\left[\frac{n!}
{\prod_{i}(n_{i}!)^{m_{i}}m_{i}!}\right]\left\{\prod_{i}
[2^{n_{i}-1}(n_{i}-1)!]^{m_{i}}({\rm tr}\,K^{n_{i}})^{m_{i}}\right\},
\end{equation}
where the sum runs over all partitions $\lambda=(n_{1})^{m_{1}}(n_{2})^{m_{2}}\cdots$
of the integer $n$.
We now have a double sum, $\sum_{n=0}^{\infty}\sum_{\lambda}$, over all
non-negative $n$ and over all partitions of $n$. But this is simply a sum over all
possible partitions of any non-negative integer. Hence, it can be re-expressed as an
unrestricted sum over each $m_{i}$, which is the number of cyclic contractions of 
$n_{i}=i$ matrices $K$.

Using $n=\sum_{i}n_{i}m_{i}$, the original sum becomes (after many cancellations)
\begin{eqnarray}
\Pi\!\!\!& =&\!\!\! \sum_{m_{1}=0}^{\infty}\sum_{m_{2}=0}^{\infty}\cdots\prod_{i}
\frac{(-1)^{n_{i}m_i}}{n_{i}^{m_{i}}m_{i}!}\left(\frac{1}{2}\right)^{m_i}
\big[\textrm{tr}\,K^{n_{i}}\big]^{m_{i}} 
=\prod_{i}\sum_{m_{i}=0}^{\infty}\frac{1}{m_i!}\left[\frac{(-1)^{i}}{2i}
\textrm {tr}\,K^{i}\right]^{m_{i}} \nonumber\\
\!\!\!& = &\!\!\! \exp\left\{-\frac{1}{2}{\rm tr}\bigg[\sum_{i}\frac{1}{i}(-1)^{(i+1)}K^{i}\bigg]
\right\}.
\end{eqnarray}
The sum is just $\ln(\mathds{1}+K)$ for the matrix argument $K$. Since for a matrix $A$,
$e^{{\rm tr}\, A}=\det e^{A}$,
\begin{equation}
\Pi=\det\exp\left[-\frac{1}{2}\ln\big(\mathds{1}+K\big)\right]=\frac{1}{\sqrt{\det(\mathds{1}+K)}}\,.
\end{equation}

\end{document}